\documentclass[journal=jpca]{achemso}
\SectionNumbersOn

\usepackage[utf8]{inputenc}
\usepackage[T1]{fontenc}
\usepackage{tgtermes}
\usepackage{indentfirst}
\usepackage{amsmath, amssymb}
\usepackage{mathtools}
\usepackage{breqn}
\usepackage{graphicx}
\usepackage{float}
\usepackage{bm}
\usepackage{physics}
\usepackage{verbatim}

\usepackage{hyperref}
\usepackage{multirow,xcolor}
\hypersetup{
	colorlinks = true,
	linkcolor = blue,
	linkbordercolor = white,
	citecolor=blue,
}
\usepackage{subfig,caption}
\newcommand{\up}{\uparrow}
\newcommand{\down}{\downarrow}

\newcommand{\rarr}{\rightarrow}

\makeatletter
\let\cat@comma@active\@empty
\makeatother

\title{Chemical Reaction Rates for Systems with Spin-Orbit Coupling and an Odd Number of Electrons: Does Berry's Phase Lead to Meaningful Spin-Dependent Nuclear Dynamics for a Two State Crossing?}
\date{\today}
\author{Yanze Wu}
\email{wuyanze@sas.upenn.edu}
\author{Gaohan Miao}
\email{gaohan@sas.upenn.edu}
\author{Joseph Subotnik}
\email{subotnik@sas.upenn.edu}
\affiliation[Department of Chemistry, University of Pennsylvania]{Department of Chemistry, University of Pennsylvania, Philadelphia, Pennsylvania 19104, USA}

\begin{document}
%\begin{tocentry}
	%\includegraphics[width=3.5cm]{toc.png}
%\end{tocentry}

\begin{abstract}
Within the context of very simple avoided crossing, we investigate the investigate the effect of a complex diabatic coupling in determining spin-dependent rate constants and scattering states. We find that, if the molecular geometry is not linear and the Berry force is not zero, one can find significant spin polarization of the products. This study emphasizes that, when analyzing nonadiabatic reactions with spin orbit coupling (and a complex Hamiltonian), one must consider how Berry force affects nuclear motion -- at least in the context of gas phase reactions. Work is currently ongoing as far as extrapolating these conclusions to the condensed phase where interesting spin selection has been observed in recent years.
\end{abstract}

\section{Introduction}
Electronic spin is one of the most fundamental observables in quantum mechanics, and manipulating spin (so-called ``spintronics'') is regarded as one of the most promising research fields.  \cite{Zutic2004,Rocha2005,Fert2008,Herrmann2010,Herrmann2011,Chumak2015,Linder2015,Jungwirth2016,Baltz2018,herrmann2018atomistic}
Over the past few decades, a plethora of techniques have shown how electronic spin can be altered and polarized through many mechanisms, including giant and tunnel magnetoresistance \cite{Baibich1988,Moodera1995}, spin-orbital torques, \cite{Gambardella2011,Brataas2012, Manchon2019} spin-transfer torques,\cite{Berger1996,Slonczewski1996,Ralph2008} spin-Hall effects \cite{Sinova2015,Hirsch1999}, and most recently (and discussed in detail below) chiral-induced spin selectivity (CISS). \cite{Naaman2012,Naaman2015,Gohler2011,Xie2011,Kettner2015,Zwang2016,Eckshtain-Levi2016,Kiran2016,Abendroth2017,Aragones2017,Bloom2017,Suda2019,Inui2020} One of the hopes of modern electronics is that, someday in the future, one will be able to produce electronic devices with low energy cost that use electronic spin for random access memories, \cite{Kent2015,Apalkov2016} sensors,\cite{Zutic2004,Parkin2003} or even quantum computers. \cite{Kane1998,Allwood2005,Chumak2015}

Now, when discussing electronic spin, there has historically been a disconnect between the chemistry and physics communities.
On the one hand, within the spintronics community in physics department, one usually supposes that the nuclei are fixed, and one investigates only how collective electronic, magnetic, and otherwise static material effects can manipulate electronic spin states.
On the other hand, within the realm of chemical dynamics, spin is usually regarded as a fixed quantum number and nuclear dynamics occur along a potential energy surface of fixed spin character (singlets or triplets, etc.).  
For the most part, chemists (who deal with nuclear dynamics and chemical reactions) and physicists (who deal with magnetic fields and electronic dynamics) have not worked on overlapping problems with synergy -- and in particular, from our point of view, the coupling between nuclear dynamics and electronic spin has not been fully explored. 

Of course, there are exceptions to the statement above. Most obviously, the electronic paramagnetic resonance (EPR) community does span both the chemistry and physics communities, and has long allowed for explicit coupling between nuclear degrees of freedom and electron spin.\cite{hoff2012advanced} Second, within the chemistry community, there has been a push as of late \cite{Mai2015,Penfold2018} to study photo-induced inter-system crossing (ISC), whereby nuclei switch between states of different spin character, as allowed by spin-orbit coupling (SOC).  
The usual picture here is that, within the realm of photochemistry, the lowest triplet state ($T_1$) is almost always lower in energy than the first excited singlet state ($S_1$),\cite{schiff1968aa} and so ISC is a very interesting {\em dynamical} (as opposed to thermodynamic) process that can precede equilibration. As such, largely led by the work of Gonzalez {\em et al},\cite{Richter2011,Richter2012,Mai2013,Richter2014,Marazzi2016,Mai2016,Sanchez-Rodriguez2017} there is now a growing desire to model the coupled nuclear-electronic-spin dynamics underlying ISC for realistic systems using variants of Tully's fewest switch surface hopping model (FSSH).\cite{Tully1990} To data, the question of how to model ISC with Tully's FSSH remains not fully solved, as recent work in our group has pointed out two subtleties regarding the nature of ISC in the context of molecules with an odd number of electrons:
\begin{itemize}
	\item When the number of electrons is odd and SOC is present, the electronic Hamiltonian becomes complex (with complex derivative couplings $\bm{d}$ \cite{Mead1979a}). Such a complex derivative coupling presents obvious difficulties for the surface hopping algorithm: how should we choose the direction to rescale nuclear momenta after a hop?\cite{Miao2019} Usually, the rescaling direction is $\bm{d}$, but if $\bm{d}$ is complex, what does this imply? Note that, if the number of electron is even and there is no magnetic field (so that there is time-reversal symmetry), the Hamiltonian and $\bm{d}$ will always be real. \cite{Mead1979a}
	\item In the presence of a complex electronic Hamiltonian, it is known that nuclei will experience a Lorentz-like ``magnetic'' force arising Berry’s phase.\cite{MichaelVictorBerry1984,Yarkony1996, Matsika2001,Matsika2001a,Matsika2002,domcke2004conical,Takatsuka2011,Subotnik2019}
\end{itemize}
Given that the complex nature of the Hamiltonian arises from the electronic spin, both of the effects represent spin-dependent nuclear phenomena.

Third, within the condensed matter physics discipline, there has also been a recent push to study spin-dependent nuclear dynamics in so far as it relates to ferromagnetic materials with strong SOC.\cite{Gambardella2011,Brataas2012,Manchon2019,Mahfouzi2016}
Here, the basic premise is that, when one runs current through a material, the spins of the transmitted electrons will interact with the built-in magnetic field of the material (which is dictated by the nuclear positions, i.e. the ``lattice'').
This interaction can lead both to changes in the spin of the transmitting electrons, as well as changes in the magnetic moment of the material (i.e. magnetic domain walls can move as nuclei relax).
%Because the conductivity of the different spins will be different in ferromagnetic materials, and there is a probe, 
%In any event,the end result is that one can use electronic current to move nuclei and order different domains (through domain wall construction and destruction). 
A recent review of this so-called ``Current-induced spin-orbit torques'' (as relevant to ferromagnetic and antiferromagnetic systems) appeared recently in Reviews of Modern Physics.\cite{Manchon2019} Overall, the question of how one couples nuclear dynamics and electronic spin appears to be in the zeitgeist nowadays.

With this background in mind, the goal of this paper is quite limited, and yet we feel prescient. In this paper, we will address a very simple question that, to our knowledge, has not yet been studied. Namely, for a simple, nonadiabatic, activated (as opposed to photo-induced) chemical reaction, how are the reaction rates and scattering cross-sections affected by the presence of electronic spin (and a complex Hamiltonian)?
We emphasize that our focus here will be on systems with two spatially distinct electronic states (or potentially two spin plus two spatially distinct electronic states), and as we will show in a future publication, the dynamics become even more interesting with three or more spatially distinct electronic states.
Nevertheless, even within a limited framework of two electronic states, we will show that Berry’s magnetic force {\em can} indeed have a significant influence on the nuclear dynamics and lead to an observable spin polarization.
%Nevertheless, it is true that these dynamical effects are always washed away in thermalized (or fully equilibrated) systems, and so in order to observe Berry's phase effects, one must either measure state-to-state transitions or operate out of equilibrium.

This paper is structured as follows: In Sec. \ref{sec:background}, we discuss some apriori well-known conditions for observing Berry’s force as induced by spin-polarization. These conditions will guide our understanding of the results below. In Sec. \ref{sec:simulation}, we present our model Hamiltonian with two spatial electronic states. In Sec. \ref{sec:results}, we present results and highlight the practical consequence of Berry's phase as far as inducing spin polarization. In Sec. \ref{sec:discussion}, we hypothesize about the potential connections between our results and exciting modern experiments that fall under the title CISS.

\section{Background} \label{sec:background}

In all that follows, unless stated otherwise, we will use $a,b$ to represent diabatic spin-free (e.g. spatial) states, $j,k$ to represent adiabatic spin-free states, $\sigma,\sigma'$ to represent spins, and $m,n$ to represent adiabatic electronic states that includes spin. 
	
\begin{figure}[H]
	\begin{center}
		\includegraphics[width=0.5\columnwidth]{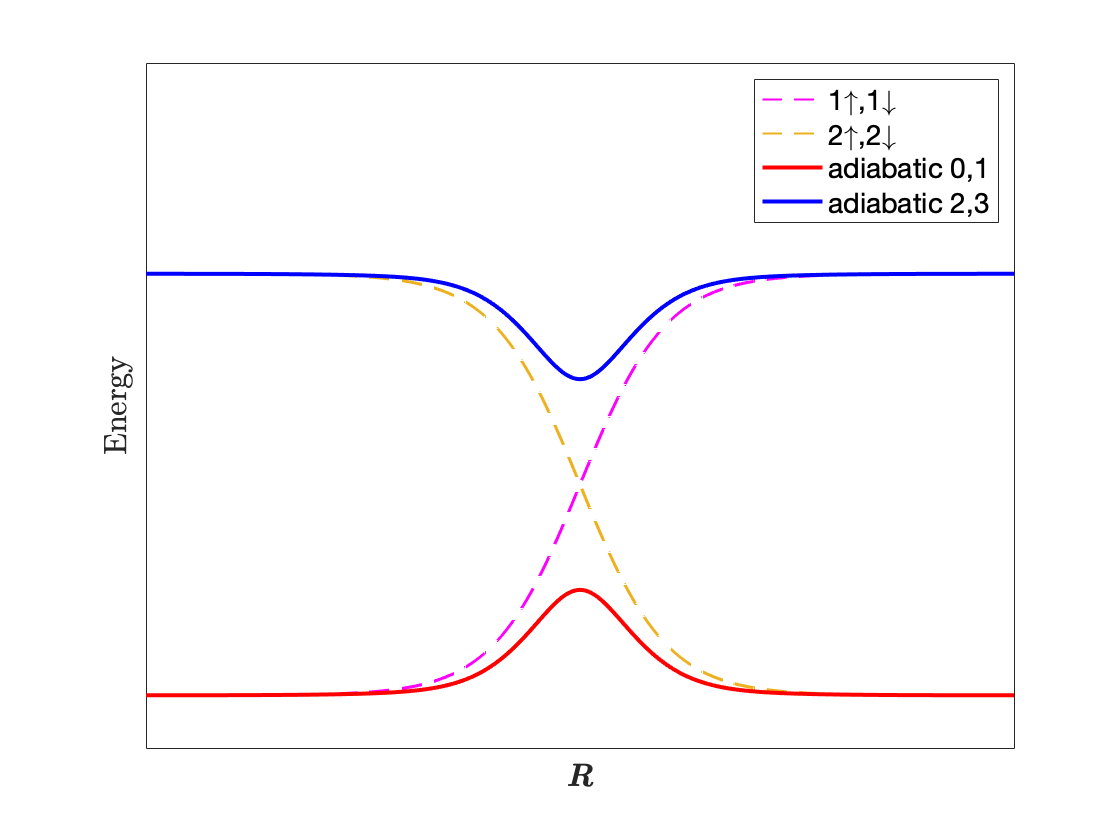}
	\end{center}
	\caption{Schematic representation of the energy surface of our model. Note that the nuclear coordinates $\bm{R}$ are two-dimensional, and here we only show the reaction coordinate.}
	\label{fig:nonadiabatic}
\end{figure}

Consider a general two-site Hamiltonian with SOC and time-reversal symmetry (TRS):
\begin{align}
   	H = -\frac{\hbar^2}{2M}\nabla^2_{\bm{R}} + \sum_{a,\sigma}{E_a(\bm{R}) c^\dagger_{a\sigma} c_{a\sigma}} + \sum_{a\ne b,\sigma}{t_{ab}(\bm{R})c^\dagger_{a\sigma}c_{b\sigma}} +
   	\sum_{a\ne b,\sigma,\sigma'}{s_{a\sigma b\sigma'}(\bm{R})c^\dagger_{a\sigma} c_{b\sigma'}}
   	\label{eq:hfull}
	\end{align}
Here, $a,b=1,2$ and $\sigma,\sigma'=\up,\down$, $M$ is the nuclear mass, $\bm{R}=(x,y)$ is the nuclear position operator, and $c_{a\sigma}$ is the electronic annihilation operator at site $a$ and spin $\sigma$. We allow the on-site energy $E_a(\bm{R})$, the diabatic coupling $t_{ab}(\bm{R})$ and the spin-orbit coupling $s_{a\sigma b\sigma'}(\bm{R})$ to vary as functions of nuclear position. By TRS, $t_{ab}(\bm{R})$ is real and $s_{a\up b\up}(\bm{R}) = s_{a\down b\down}^*(\bm{R})$ and $s_{a\up b\down}(\bm{R}) = -s_{a\down b\up}^*(\bm{R})$. As shown in Fig. \ref{fig:nonadiabatic}, we consider the case where there is a curve crossing between the two states, which results in an energy barrier in the middle. Both the adiabatic energy surfaces and diabatic energy surfaces are doubly degenerate because of Kramer's rule.
	
Mathematically, within the basis $\ket{1\up}, \ket{1\down}, \ket{2\up}, \ket{2\down}$, the Hamiltonian can be written as follows:
\begin{align}
	H = -\frac{\hbar^2}{2M}\nabla^2_{\bm{R}} + \begin{bmatrix}
    	E_1(\bm{R}) & 0 & V_t(\bm{R}) & V_s(\bm{R}) \\
    	0 & E_1(\bm{R}) & -V_s^*(\bm{R}) & V_t^*(\bm{R}) \\
    	V_t^*(\bm{R}) & -V_s(\bm{R}) & E_2(\bm{R}) & 0 \\
    	V_s^*(\bm{R}) & V_t(\bm{R}) & 0 & E_2(\bm{R})
	\end{bmatrix}
	\label{eq:h4state}
\end{align}
where $V_t = t + s_{1\up 2\up}$ and $V_s = s_{1\up 2\down}$. Note that the adiabatic ground state energy (doubly degenerate) of Hamiltonian \eqref{eq:h4state} is simply
\begin{align}
   	E_g(\bm{R}) = \frac{E_1(\bm{R})+E_2(\bm{R})}2 - \sqrt{{\Big(\frac{E_1(\bm{R})-E_2(\bm{R})}2\Big)}^2 + V_t^2(\bm{R}) + V_s^2(\bm{R})}
   	\label{eq:ground}
\end{align}
which can be analytically derived via matrix diagonalization.
	
\subsection{Berry Curvature of Complex Hamiltonians} \label{sec:berry}
	
It is important to remember that, when an electronic Hamiltonian is not real (i.e. $\Im{H}\ne 0$, as in Hamiltonian \eqref{eq:h4state}), dynamics along the ground state do not follow the usual (conservative) Newton's laws: The forces that nuclei feel are not just the simple energy gradient $-\nabla E_g$ (where $E_g$ is defined in Eq.~\eqref{eq:ground}).
In particular, for an arbitrary system (with or without TRS) moving along adiabatic surface $m$, nuclei feel an effective ``Berry magnetic force'', defined by\cite{MichaelVictorBerry1984, Mead1979, Subotnik2019}:
\begin{align} \label{eq:berryforce}
   	\bm{F}^{B}_m = \hbar \frac{\bm{p}}M\times \bm{B}_m = \frac{2\hbar}{M} \Im{\sum_{n\ne m}{\bm{d}_{mn}(\bm{p}\cdot \bm{d}_{nm})}}
\end{align}
Here, $\bm{d}_{mn}$ is the derivative coupling between surface $m$ and $n$:
\begin{align} \label{eq:drvcoupling}
    \bm{d}_{mn} = \frac{\mel{\psi_m}{\nabla H}{\psi_n}}{E_n - E_m}
\end{align}
In 2D and 3D systems, this Berry force is equivalent to the Lorentz force produced by a magnetic field of the form:
\begin{align} \label{eq:bfield}
   	\bm{B}_m = \nabla \times (i\ev{\nabla}{\psi_m}) = -i\sum_{n\ne m}{\bm{d}_{mn}\times \bm{d}_{nm}}
\end{align}

If a general Hamiltonian does not obey TRS, $\bm{F}^B_m$ can be different for each of the adiabats, such that nuclei can in principle experience completely different magnetic forces on different surfaces. That being said, when a system does have TRS and an odd number of electrons, from Kramer's rule, each adiabatic surface will be doubly degenerate with a time reversal state (``Kramer doublets''), and the Berry forces on the two surfaces are related to each other. To prove this statement, suppose states $m'$, $n'$ are the time-reversal states of $m$, $n$, i.e. $\ket{m'} = \Theta\ket{m}$ and  $\ket{n'} = \Theta\ket{n}$, where $\Theta$ is the time-reversal operator. The derivative coupling shown in Eq.~\eqref{eq:drvcoupling} between $m'$ and $n'$ is
\begin{align} 
	\bm{d}_{m'n'} &= \frac{\mel{\psi_{m'}}{\nabla H}{\psi_{n'}}}{E_{n'} - E_{m'}}
	= \frac{\mel{\Theta\psi_{m}}{(\nabla H)\Theta}{\psi_{n}}}{E_{n} - E_{m}}
\end{align}
Because $H$ obeys TRS, we have $H\Theta = \Theta H$ (and hence $(\nabla H)\Theta = \Theta (\nabla H)$), and thus
\begin{align}
    \bm{d}_{m'n'} = \frac{\braket{\Theta \psi_m}{\Theta (\nabla H)\psi_n} }{E_n - E_m}
\end{align}
By the antiunitarity of $\Theta$, $\braket{\Theta u}{\Theta v} = {\braket{u}{v}}^*$, so that
\begin{align} \label{eq:drvcouplingtrs} 
	\bm{d}_{m'n'} = \frac{{\mel{\psi_{m}}{\nabla H}{\psi_{n}}}^*}{E_{n} - E_{m}} = \bm{d}_{mn}^*
\end{align}
The corresponding Berry magnetic force on adiabatic surface $m'$ is
\begin{align} \label{eq:berryforcetrs}
    \bm{F}^{B}_{m'} = \frac{2\hbar}{M} \Im{\sum_{n'\ne m'}{\bm{d}_{m'n'}(\bm{p}\cdot \bm{d}_{n'm'})}}
     = \frac{2\hbar}{M} \Im{\sum_{n\ne m}{\bm{d}_{mn}^*(\bm{p}\cdot \bm{d}_{nm}^*)}}
     = -\bm{F}^{B}_m
\end{align}
Eq.~\eqref{eq:berryforcetrs} shows that, given TRS and the same momentum, the nuclei feel equal and opposite forces on different Kramer doublet surfaces.

\subsection{Detailed Balance and Equilibrium For Systems with TRS} \label{sec:balance}

Although our primary interest is the nature of chemical reactions {\em in the condensed phase}, for the present paper, we will simulate below the small unbound system shaped as in Fig. \ref{fig:nonadiabatic} within a microcanonical ensemble. In restricting ourselves to a small system, our rationale is that, because we can employ exact scattering theory for small systems, we will be able to avoid any and all semiclassical approximations and make definitive statements about quantum dynamics without regard to any quantities of accuracy; all results here should be appropriately extendable to larger, open systems with some caveats. See Sec. \ref{sec:discusscondensed} for a more detailed discussion of this point.
With this framework in mind, consider a system whereby two molecules can undergo a binary collision (with both electrons and nuclei rearranging), and we will label the reactant terminal as terminal \#1 and the product terminal as terminal \#2. For each terminal, we label the interior quantized states as $\mu$/$\nu$ (terminal \#1) or $\mu'$/$\nu'$ (terminal \#2); $\sigma,\sigma'=\up,\down$ denote specific spin states.
We assume that there is no SOC asymptotically within terminals \#1 or \#2, so that one can label all incoming and outgoing states with pairs of labels $\mu\sigma$, etc.

The very first question we must ask is: What is the nature of equilibrium for such a system? After a quick second of thought, one realizes that there cannot be any spin polarization in the terminals. After all, at equilibrium, we must find that the population of electronic state $\mu$ with spin $\up$ in terminal \#1 is:
\begin{align}
    P_{1,\mu\up} &= \frac{1}{Z}\mel{1,\mu\up}{e^{-\beta H}}{1,\mu\up} 
    = \frac{1}{Z}\mel{1,\mu\up}{e^{-\beta \Theta^{-1} H \Theta}}{1,\mu\up} \nonumber \\
    &= \frac{1}{Z}\mel{1,\mu\up}{\Theta^{-1} e^{-\beta  H }\Theta}{1,\mu\up}
\end{align}
where $Z$ is the partition function.
Using the fact that $\Theta = KU$, where $K$ is the antilinear complex conjugation operator and $U$ is unitary operator that interconverts between up and down spin, it follows that:
\begin{align}
    P_{1,\mu \up} &= \frac{1}{Z}\mel{1,\mu\down}{K^{-1} e^{-\beta  H }K}{1,\mu\down} 
    = \frac{1}{Z}\mel{1,\mu\down}{e^{-\beta  H^* }}{1,\mu\down} \nonumber \\
    &= \frac{1}{Z}\mel{(1,\mu)^*\down}{e^{-\beta  H } }{(1,\mu)^*\down}
\end{align}
Using the fact that there is no SOC asymptotically, one can assume that the spatial state $\ket{1,\mu}$ has a real wave function, and so at equilibrium:
\begin{align} \label{eq:balancespin}
    P_{1,\mu\up} = P_{1,\mu\down}
\end{align}

For our purposes below, it will be helpful to rederive these equilibrium (Boltzmann) statistics for the internal quantum states explicitly using scattering theory. 
According to time reversibility, we know that for a scattering process with translational energy $E$, the transmission rate $T_{a,\mu\sigma\rarr b\mu'\sigma'}$ satisfies \cite{Zhai2005}
\begin{align} \label{eq:balance1}
	T_{a,\mu\sigma\rarr b,\mu'\sigma'}(E) &= T_{b,\mu'\tilde{\sigma}'\rarr a,\mu\tilde{\sigma}}(E+\epsilon_\mu-\epsilon_{\mu'})
\end{align}
Here, $a,b$ are indices of terminals (either the same or different), $\epsilon_\mu,\epsilon_{\mu'}$ are the bound energies of the quantized states $\mu,\mu'$ and $\tilde{\sigma},\tilde{\sigma}'$ are the opposite spin of $\sigma,\sigma'$. In other words, if $\sigma=\up$, then $\tilde{\sigma}=\down$, etc.

Now, we suppose that the terminals are in contact with a bath at temperature $kT_{bath} =  1/\beta$ that promotes collisions and dictates (on average) the translational energy of collision as $E$. 
Let us define the state-to-state rate constant
\begin{align} \label{eq:rc}
    k_{a,\mu\sigma\rarr b,\mu'\sigma'} &= \frac{1}{\beta}\int{dEe^{-\beta E}T_{a,\mu\sigma\rarr b,\mu'\sigma'}(E)}
\end{align}
In terms of $k$, Eq.~\eqref{eq:balance1} reads
\begin{align} \label{eq:balance1_2}
    k_{a,\mu\sigma\rarr b,\mu'\sigma'} &= e^{\epsilon_\mu-\epsilon_{\mu'}}k_{b,\mu'\tilde{\sigma}'\rarr a,\mu\tilde{\sigma}}
\end{align}

There are two key properties of these state-to-state resolved transmission factors and rate constants that must be emphasized.
First, in Eq.~\eqref{eq:balance1}, by letting $b=a$ and $\mu'=\mu$, we recover the equality
\begin{align} \label{eq:balancespecial0}
    T_{a,\mu\sigma\rarr a,\mu\sigma}(E) = T_{a,\mu\tilde{\sigma}\rarr a,\mu\tilde{\sigma}}(E)
\end{align}
i.e. the reflection probability is not changed by switching spin.

Second, according to the conservation of probability,
\begin{align}
    \sum_{b,\mu'\sigma'}{T_{a,\mu\sigma\rarr b,\mu'\sigma'}(E)}=\sum_{b,\mu'\sigma'}{T_{a,\mu\tilde{\sigma}\rarr b,\mu'\sigma'}(E)}=1
\end{align}
Therefore, if we invoke Eq.~\eqref{eq:balancespecial0}, we find:
\begin{align} \label{eq:balancespecial1}
    \sum_{\substack{b,\mu'\sigma \\ (b,\mu'\sigma'\ne a,\mu\sigma)}}{T_{a,\mu\sigma\rarr b,\mu'\sigma'}(E)} = \sum_{\substack{b,\mu'\sigma \\ (b,\mu'\sigma'\ne a,\mu\tilde{\sigma})}}{T_{a,\mu\tilde{\sigma}\rarr b,\mu'\sigma'}(E)}
\end{align}
Or, in terms of $k$,
\begin{align} \label{eq:balancespecial2}
    \sum_{\substack{b,\mu'\sigma \\ (b,\mu'\sigma'\ne a,\mu\sigma)}}{k_{a,\mu\sigma\rarr b,\mu'\sigma'}} = \sum_{\substack{b,\mu'\sigma \\ (b,\mu'\sigma'\ne a,\mu\tilde{\sigma})}}{k_{a,\mu\tilde{\sigma}\rarr b,\mu'\sigma'}}
\end{align}

At this point, we can prove that Boltzmann statistics arise. According to standard kinetics, the change in population of any one internal state is:
\begin{align} \label{eq:dpdt}
	 \dv{P_{a,\mu\sigma}}{t} = \sum_{\substack{b,\mu'\sigma \\ (b,\mu'\sigma'\ne a,\mu\sigma)}}
     {(-P_{a,\mu\sigma}k_{a,\mu\sigma\rarr b,\mu'\sigma'} + P_{b,\mu'\sigma'}k_{b,\mu'\sigma'\rarr a,\mu\sigma})}
\end{align}
If we apply Eq.~\eqref{eq:balancespecial2} to the first term of Eq.~\eqref{eq:dpdt} and we apply Eq.~\eqref{eq:balance1_2} to the second term of Eq.~\eqref{eq:dpdt}, we arrive at:
\begin{align} 
	 \dv{P_{a,\mu\sigma}}{t} &= 
     \sum_{\substack{b,\mu'\sigma \\ (b,\mu'\sigma'\ne a,\mu\tilde{\sigma})}}
     {(-P_{a,\mu\sigma}k_{a,\mu\tilde{\sigma}\rarr b,\mu'\sigma'} + e^{\beta(\epsilon_{\mu'}-\epsilon_\mu)}P_{b,\mu'\sigma'}k_{a,\mu\tilde{\sigma}\rarr b,\mu'\sigma'})} \nonumber \\
     &= \sum_{\substack{b,\mu'\sigma \\ (b,\mu'\sigma'\ne a,\mu\tilde{\sigma})}}
     {e^{-\beta\epsilon_\mu}k_{a,\mu\tilde{\sigma}\rarr b,\mu'\sigma'}(-e^{\beta\epsilon_\mu}P_{a,\mu\sigma} + e^{\beta\epsilon_{\mu'}}P_{b,\mu'\sigma'})}
\end{align}
Similar expressions can be obtained starting from $\dv{P_{b,\mu'\sigma'}}{t}$. Ultimately, because all populations must be stationary at equilibrium ($\dv{P_{a,\mu\sigma}}{t} = 0$), we find that the equilibrium populations must indeed satisfy:
\begin{align} 
    P_{a,\mu\sigma} = \frac{e^{-\beta \epsilon_\mu}}{Z}
\end{align}
Here $Z=\sum_{\mu}{e^{-\beta\epsilon_\mu}}$ is a normalization constant, i.e. the total partition function.

For the seasoned reader, the above arguments just recapitulate facts that are well known from statistical mechanics:
detailed balance at equilibrium follows from the time-reversible dynamics of a system in contact with a thermal bath.
And as long as SOC vanishes far away from a crossing point, so that spin is a good quantum number in the terminals,
Boltzmann equilibrium statistics still follow from the time-reversal symmetry properties of the transmission probabilities (despite the more complicated spin-dependent nature of those probabilities in Eq.~\eqref{eq:balance1}).
%Thus, in order to study dynamics and the approach to equilibrium, one can gain a lot of information working with the
%spin-dependent transmission amplitudes (as we do below).
Nevertheless, the argument above tells us only about equilibrium, and not the {\em approach} to equilibrium.
With this goal in mind, let us now turn our attention to reaction rates and dynamics, using the transmission probabilities defined above.

\subsection{Reaction Rates for Systems with TRS} \label{sec:rates}
Given the background above, let us consider a simple thought experiment.
Suppose we start with a system out of equilibrium, or more specifically a system in {\em local equilibrium only}, i.e. with all population in one terminal. Given TRS, from the argument above, we will necessarily find that -- at long times -- the systems will reach an equilibrium state without any spin-polarized populations anywhere.
Nevertheless, one can ask: what about short times? Will one ever observe spin-polarized populations at short times? 
Or vice versa, in language adapted to the canonical ensemble, for a system with time reversal symmetry, can we find spin-dependent reaction rate constants?

In general, yes, we can find spin-dependent rate constants. Or, in blunt language, suppose that, at the beginning of a simulation, the system is equilibrated only at terminal \#1, i.e. $P_{1,\mu\up} = P_{1,\mu\down} = \frac{e^{-\beta\epsilon_\mu}}{Z_1}$ (where $Z_1$ is the partition function of terminal \#1).
For very short times, we can ignore the build-up of population in other terminals or any changes in $P_{1,\mu\up},P_{1,\mu\down}$ as functions of time. Therefore, the short time expression for polarization in any other terminal (say \#2) becomes
\begin{align} \label{eq:defpol}
	k_{polar} &= \sum_{\mu'}{\dv{P_{2,\mu'\up} - P_{2,\mu'\down}}{t}} \nonumber \\
	&= \sum_{\mu,\mu'}{P_{1,\mu\up}(k_{1,\mu\up\rarr 2,\mu'\up} - k_{1,\mu\up\rarr 2,\mu'\down}) + P_{1,\mu\down}(k_{1,\mu\down\rarr 2,\mu'\up} - k_{1,\mu\down\rarr 2,\mu'\down})} \nonumber \\
    &= \frac{1}{Z_1}\sum_{\mu,\mu'}{e^{-\beta\epsilon_\mu}(k_{1,\mu\up\rarr 2,\mu'\up} - k_{1,\mu\up\rarr 2,\mu'\down} + k_{1,\mu\down\rarr 2,\mu'\up} - k_{1,\mu\down\rarr 2,\mu'\down})}
\end{align}
In general, $k_{1,\mu\up\rarr 2,\mu'\up} \ne k_{1,\mu\down\rarr 2,\mu'\down}$ and $k_{1,\mu\down\rarr 2,\mu'\up} \ne k_{1,\mu\up\rarr 2,\mu'\down}$, so the polarization in terminal \#2 (as given by Eq.~\eqref{eq:defpol}) is not zero. Our goal in this paper will be to analyze the size of these spin polarizations.

That being said, there are two caveats to this general conclusion that must now be discussed; in particular, there are two cases where spin polarization will never occur.

\subsubsection{Caveat \#1: System with Inversion Symmetry} \label{sec:inversion}

\begin{comment}
For a time-reversal symmetric Hamiltonian, the transmission of different spin are connected by (where the terminals $a,b$ can be the same or different)
\begin{align}
	\begin{cases}
		T_{a,\mu\up\rarr b,\mu'\up}(E) = T_{b,\mu'\down\rarr a,\mu\down}(E+\epsilon_\mu-\epsilon_{\mu'}) \\
		T_{a,\mu\down\rarr b,\mu'\down}(E) = T_{b,\mu'\up\rarr a,\mu\up}(E+\epsilon_\mu-\epsilon_{\mu'}) \\
		T_{a,\mu\up\rarr b,\mu'\down}(E) = T_{b,\mu'\up\rarr a,\mu\down}(E+\epsilon_\mu-\epsilon_{\mu'}) \\
		T_{a,\mu\down\rarr b,\mu'\up}(E) = T_{b,\mu'\down\rarr a,\mu\up}(E+\epsilon_\mu-\epsilon_{\mu'}) 
	\end{cases}
	\label{eq:trsspin0}
\end{align}
Substituting Eq.~\eqref{eq:trsspin0} into Eq.~\eqref{eq:rc} gives
\begin{align}
	\begin{cases}
		k_{a,\mu\up\rarr b,\mu'\up} = k_{b,\mu'\down\rarr a,\mu\down} \\
		k_{a,\mu\down\rarr b,\mu'\down} = k_{b,\mu'\up\rarr a,\mu\up} \\
		k_{a,\mu\up\rarr b,\mu'\down} = k_{b,\mu'\up\rarr a,\mu\down} \\
		k_{a,\mu\down\rarr b,\mu'\up} = k_{b,\mu'\down\rarr a,\mu\up} 
	\end{cases}
	\label{eq:trsspin}
\end{align}
where $a,b$ are indices of terminals.
\end{comment}

For a system with time reversibility, the transmission rates are connected by Eq.~\eqref{eq:balance1}. Moreover, if the system also satisfies inversion symmetry (IS) between terminal \#1 and \#2, then for inversion symmetric pairs of states $\ket{1,\mu}$, $\ket{2,\tilde{\mu}}$ and $\ket{1,\nu}$, $\ket{2,\tilde{\nu}}$, we have $\epsilon_\mu=\epsilon_{\tilde{\mu}}$, $\epsilon_\nu=\epsilon_{\tilde{\nu}}$, and
\begin{align} \label{eq:invsym}
	k_{1,\mu\up\rarr 2,\tilde{\nu}\up} = k_{2,\tilde{\mu}\up \rarr 1,\nu\up} \nonumber \\
	k_{1,\mu\up\rarr 2,\tilde{\nu}\down} = k_{2,\tilde{\mu}\up \rarr 1,\nu\down}
\end{align}
In such a case, if the system begins in equilibrium in terminal \#1, then the overall spin polarization in terminal \#2 is zero even at short times:
\begin{align} \label{eq:invsym2}
	k_{polar} &= \frac{1}{Z_1}\sum_{\mu,\mu'}{e^{-\beta\epsilon_\mu}(k_{1,\mu\up\rarr 2,\mu'\up} + k_{1,\mu\down\rarr 2,\mu'\up} - k_{1,\mu\down\rarr 2,\mu'\down} - k_{1,\mu\up\rarr 2,\mu'\down})} \nonumber \\
	&= \frac{1}{Z_1}\sum_{\mu,\mu'}{e^{-\beta\epsilon_\mu}(k_{1,\mu\up\rarr 2,\mu'\up} + k_{1,\mu\down\rarr 2,\mu'\up} - e^{\beta(\epsilon_\mu-\epsilon_{\mu'})}(k_{2,\mu'\up\rarr 1,\mu\up} + k_{2,\mu'\down\rarr 1,\mu\up}))} \text{ (by TRS)} \nonumber \\
    &= \frac{1}{Z_1}\sum_{\mu,\nu}{e^{-\beta\epsilon_\mu}(k_{1,\mu\up\rarr 2,\tilde{\nu}\up} + k_{1,\mu\down\rarr 2,\tilde{\nu}\up})}
    - \frac{1}{Z_1}\sum_{\mu,\nu}{e^{-\beta\epsilon_{\tilde{\mu}}}(k_{2,\tilde{\mu}\up\rarr 1,\nu\up} + k_{2,\tilde{\mu}\down\rarr 1,\nu\up})} \text{ (by IS)} \nonumber \\    
	&= 0  \text{ (by IS)}
\end{align}
The significance of Eq.~\eqref{eq:invsym2} is that, for a system with inversion symmetry,
provided that one equilibrates all dynamics initially in one terminal, no spin polarization will ever be observed in the other terminal. 
There is no such constraint for system lacking inversion symmetry, such as system with a helical potential.\cite{Yeganeh2009,Eremko2013,Gutierrez2012}

\subsubsection{Caveat \#2: System with Two Terminals and No Spin Flip} \label{sec:twoterminal}

Although our entire discussion above has been in the context of a system with two terminals, all of the conclusions above were in fact general and applicable to the case of arbitrarily many terminals. However, for a system with two terminals (and not three or more) and no possibility of spin flip, there is one more key symmetry worth mentioning. 
Namely, if we assume that one terminal has an equilibrated set of internal quantum states, the total transmission will necessarily be equal for the up spin and down spin; thus, no spin polarization can be observed.
%and we calculate the transmission coefficient 
%from one terminal to the other terminal (which is akin to the total rate constant).

To prove this statement, we will begin with most general Green's function expression for transmission (valid for the case of arbitrarily many terminals):\cite{nitzan2006chemical} 
\begin{align} \label{eq:green1}
	& T_{1,\mu\sigma\rarr 2,\mu'\sigma'}(E) = \Tr[\Gamma_{1,\mu\sigma}(E+\epsilon_\mu) G^\dagger(E+\epsilon_\mu)\Gamma_{2,\mu'\sigma'}(E+\epsilon_\mu) G(E+\epsilon_\mu)] \\
	& G(E+\epsilon_\mu) = \frac{1}{E +\epsilon_\mu - H - \Sigma(E+\epsilon_\mu)}
\end{align}
where $\Sigma$ is the self-energy and $\Gamma_{1,\mu\sigma},\Gamma_{2,\mu\sigma}$ are the imaginary parts of the self-energies for terminals \#1 and \#2 with spin $\sigma$ and state $\mu$.

%Second, when all the substates are divided into two categories, there is a reciprocal relation between the overall rates \cite{Datta1990}:
%    \begin{align} \label{eq:twoterm}
%        \sum_{(\epsilon,\mu)\in N, (\epsilon',\nu)\notin N}{T_{\epsilon,\mu\rarr \epsilon',\nu}(E)} = \sum_{(\epsilon,\mu)\in N, (\epsilon',\nu)\notin N}{T_{\epsilon',\nu\rarr \epsilon,\mu}(E)}
%    \end{align}
%where $N$ is an arbitrary collection of states.
Since there is no spin-flipping transmission, the spin polarization rate $k_{polar}$ defined in Eq.~\eqref{eq:defpol} is just
\begin{align} \label{eq:spinrc0}
    k_{polar} &= \frac{1}{\beta Z_1}\int{dE\sum_{\mu,\mu'}{e^{-\beta(E+\epsilon_\mu)}(T_{1,\mu\up\rarr 2,\mu'\up}(E) - T_{1,\mu\down\rarr 2,\mu'\down}(E))}}
\end{align}
By time-reversibility (Eq.~\eqref{eq:balance1}), we have
\begin{align} \label{eq:spinrc1}
    k_{polar} &= \frac{1}{\beta Z_1}\int{dE\sum_{\mu,\mu'}{e^{-\beta(E+\epsilon_\mu)}(T_{1,\mu\up\rarr 2,\mu'\up}(E) - T_{2,\mu\up\rarr 1,\mu'\up}(E))}} \nonumber \\
    &= \frac{1}{\beta Z_1}\int{dE_{tot} e^{-\beta E_{tot}} \Big(\Tr[
        \Big(\sum_{\mu}{\Gamma_{1,\mu\up}(E_{tot})}\Big) G^\dagger(E_{tot}) \Big(\sum_{\mu'}{\Gamma_{2,\mu'\up}(E_{tot})}\Big) G(E_{tot})]} \nonumber \\
    &\quad - \Tr[\Big(\sum_{\mu}{\Gamma_{2,\mu\up}(E_{tot})}\Big) G^\dagger(E_{tot}) \Big(\sum_{\mu'}{\Gamma_{1,\mu'\up}(E_{tot})}\Big) G(E_{tot})]\Big)
\end{align}

Now, most generally, the overall Hamiltonian can be divided into blocks:
\begin{align}
    H = \begin{bmatrix} H_{\up\up} & H_{\up\down} \\ H_{\down\up} & H_{\down\down} \end{bmatrix}
\end{align}
When there is no spin flip (i.e. $H_{\up\down} = H_{\down\up} = 0$), $H_{\up\up}$ and $H_{\down\down}$ are independent blocks. For the up spin, only $H_{\up\up}$ enters into the expression. By defining $G_\up(E) = (E-H_{\up\up}-\Sigma_\up)^{-1}$, $\Gamma_{1\up} = \sum_\mu{\Gamma_{1,\mu\up}}$ and $\Gamma_{2\up} = \sum_{\mu'}{\Gamma_{2,\mu'\up}}$, Eq.~\eqref{eq:spinrc1} can be simplified as (for 
\begin{align} \label{eq:spinrc}
    k_{polar} &= \frac{1}{\beta Z_1}\int{dE e^{-\beta E}\Big( \Tr[\Gamma_{1\up}G^\dagger_\up\Gamma_{2\up}G_\up] - \Tr[\Gamma_{2\up}G^\dagger_\up\Gamma_{1\up}G_\up]\Big)}
\end{align}
For notational convenience, we omit the energy dependence in $\Gamma$ and $G$ in Eq.~\eqref{eq:spinrc}-\eqref{eq:rateequ2}, but they remain functions of energy. To prove $k_{polar}=0$, let us show that
\begin{align} \label{eq:twotermsym}
	\Tr[\Gamma_{1\up}G^\dagger_\up\Gamma_{2\up}G_\up]=\Tr[\Gamma_{2\up}G^\dagger_\up\Gamma_{1\up}G_\up]
\end{align}
%This result is general: the rate of transfer from $1\rarr 2$ equal the rate of transferring from $2\rarr 1$ (Eq.~\eqref{eq:twotermsym} holds) for any system with two terminals with or without time reversal symmetry.
%And equally, if we return to the two-terminal case in Eq.~\eqref{eq:twotermsym} above, we can invoke the time-reversal symmetric relation in Eq.~\eqref{eq:balance1_2}, $k_{1\down\rarr 2\down} = k_{2\up\rarr 1\up}$, to prove that no spin polarization will ever emerge if we start in terminal \#1 at local equilibrium.
By separating the real and imaginary part of $\Sigma_\up$, $\Sigma_\up = \Lambda_\up + i\Gamma_\up$, and define $\Omega_{\up\up} = E - H_{\up\up} - \Lambda_\up$., the LHS of Eq.~\eqref{eq:twotermsym} can be expanded as
\begin{align} \label{eq:twotermderive1}
    \Tr[\Gamma_{1\up}G^\dagger_\up\Gamma_{2\up}G_\up] &= \Tr[\Gamma_{1\up}\frac{1}{E - H_{\up\up} - \Sigma_\up^\dagger}\Gamma_{2\up}\frac{1}{E - H_{\up\up} - \Sigma_\up}] \nonumber \\
    &= \Tr[\Gamma_{1\up}\frac{1}{\Omega_{\up\up} + i\Gamma_\up}\Gamma_{2\up}\frac{1}{\Omega_{\up\up} - i\Gamma_\up}]
\end{align}
The key point is that, when there are {\em only two terminals}, $\Gamma_\up = \Gamma_{1\up}+\Gamma_{2\up}$. Therefore,
\begin{align} \label{eq:twotermderive2}
    \Tr[\Gamma_{1\up}G^\dagger_\up\Gamma_{2\up}G_\up] &+ \Tr[\Gamma_{2\up}\frac{1}{\Omega_{\up\up}+i\Gamma_\up}\Gamma_{2\up}\frac{1}{\Omega_{\up\up}-i\Gamma_\up}] \nonumber \\
    &=\Tr[\Gamma_{\up}\frac{1}{\Omega_{\up\up}+i\Gamma_\up}\Gamma_{2\up}\frac{1}{\Omega_{\up\up}-i\Gamma_\up}] 
\end{align}
Vice versa, it is also true that:
%\nonumber \\
%    &\quad\quad= \Tr[\Gamma_{\up}\frac{1}{\Omega_{\up\up}+i\Gamma_\up}\Gamma_{2\up}\frac{1}{\Omega_{\up\up}-i\Gamma_\up}] \nonumber \\
%    &\quad\quad= -\frac{i}{2} \Tr[\frac{1}{\Omega_{\up\up}-i\Gamma_\up}2i\Gamma_{\up}\frac{1}{\Omega_{\up\up}+i\Gamma_\up}\Gamma_{2\up}] \nonumber \\
%    &\quad\quad= -\frac{i}{2} \Tr[\Big(\frac{1}{\Omega_{\up\up}-i\Gamma_\up}-\frac{1}{\Omega_{\up\up}+i\Gamma_\up}\Big)\Gamma_{2\up}] \nonumber \\
%    &\quad\quad= \frac{i}{2} \Tr[\frac{1}{\Omega_{\up\up}+i\Gamma_\up}(-2i\Gamma_{\up})\frac{1}{\Omega_{\up\up}-i\Gamma_\up}\Gamma_{2\up}] \nonumber \\  
%    \nonumber \\ 
\begin{align} \label{eq:twostatesymf}
    \Tr[\Gamma_{2\up}G^\dagger_\up\Gamma_{1\up}G_\up] &+ \Tr[\Gamma_{2\up}\frac{1}{\Omega_{\up\up}+i\Gamma_\up}\Gamma_{2\up}\frac{1}{\Omega_{\up\up}-i\Gamma_\up}] \nonumber \\
    &=\Tr[\Gamma_{2\up}\frac{1}{\Omega_{\up\up}+i\Gamma_\up}\Gamma_\up\frac{1}{\Omega_{\up\up}-i\Gamma_\up}]
\end{align}
Lastly, using the well known identity $D^{-1} - C^{-1} = D^{-1}(D - C)^{-1}C^{-1}$ and setting $D = \Omega_{\up\up} - i\Gamma_\up$ and $C = \Omega_{\up\up}+i\Gamma_\up$, it follows that Eq.~\eqref{eq:twotermsym} holds. Therefore, if we begin in equilibrium in terminal \#1, we can never find any polarization in terminal \#2.

Furthermore, according to Eq.~\eqref{eq:twotermsym}, for a system with only two terminals (with or without time reversal symmetry), the rate constant must satisfy $k_{1\rarr 2}Z_1 = k_{2\rarr 1}Z_2$, as follows from the general expressions:
\begin{align} 
    k_{1\rarr 2} = \frac{1}{\beta Z_1}\int{dE e^{-\beta E} \Tr[\Gamma_{1}G^\dagger\Gamma_{2}G]} \\
    k_{2\rarr 1} = \frac{1}{\beta Z_2}\int{dE e^{-\beta E} \Tr[\Gamma_{2}G^\dagger\Gamma_{1}G]} \label{eq:rateequ2}
\end{align}
We will address this result in Sec. \ref{sec:discussionmultiple} in more detail.

%And equally, if we return to the two-terminal case in Eq.~\eqref{eq:twotermsym} above, we can invoke the time-reversal symmetric relation in Eq.~\eqref{eq:balance1_2}, $k_{1\down\rarr 2\down} = k_{2\up\rarr 1\up}$, to prove that no spin polarization will ever emerge if we start in terminal \#1 at local equilibrium. Eq.~\eqref{eq:}

This concludes the necessary background as far as understanding spin-dependent rate constants.

\section{Simulation Details} \label{sec:simulation}

\subsection{Observables}
In order to gain information for spin-dependent reaction rates, below we will study a quasi-1D system where the diabatic surfaces $E_1$ and $E_2$ are unbound in one direction. We imagine a simple complex avoided crossing (see Fig. \ref{figure:geometry}), and we define two coordinates: (1) a longitudinal (reaction) coordinate, where nuclei are incoming and outgoing, and along which the magnitude of the potential energy surfaces changes; (2) a transverse coordinate, where the nuclear wave function is bound and along which only the phase of the diabatic coupling changes.
The nuclei will be given a low enough energy such that one always starts and ends on the ground adiabatic electronic state. Nevertheless, in the transverse coordinate, there is enough energy such that the nuclei can switch between different bound states. Therefore, each individual state of interest is characterized by two numbers: the transverse eigenstate $n_T$ and electronic spin $\sigma$.
By design, all SOC parameters are negligible outside the reaction region, so that the spin is a good quantum number for the incoming and outgoing waves.

From the arguments in the previous section, it is clear that, for the case of two terminals, many spin-dependent effects can be washed away if we average over an equilibrated set of initial conditions. With this caveat in mind, below, we will report state-to-state transmission (rather than a total rate constant that would be the integral over many state-to-state transmission rates).   In particular, in what follows below, we will report transmission rates from the ground transverse state ($n_T = 0$) of electronic diabatic state 1 to all other possible transverse states corresponding to diabatic state 2.
In other words, if the initial spin is $\sigma$ and we denote the final spin $\sigma'$, the spin-dependent transmission rate we report is:
\begin{align} \label{eq:transmission}
	T_{\sigma\rarr\sigma'} = \sum_{n_T}{T_{1,0\sigma\rarr 2,n_T\sigma'}}
\end{align}
% An analysis of the {\em total}, temperature-dependent rate constants under different conditions (e.g. nonequilbrium initial conditions, simulations with more than two terminals, etc.} will be presented in an upcoming paper.

%Forces are not easily observable, and Berry magnetic force is no different. Nevertheless, a magnetic force can certainly affect dynamics. However, from the argument in \ref{sec:rates}, in order to observe Berry phase in a system with TRS, we must investigate the individual state-to-state reaction rates. In this paper, in order to have well-defined individual states and reaction rates, 
%Because spin is a good quantum number for incoming and outgoing states, the Kramer doublets for these states have the same spatial part but opposite spins. As discussed in Section \ref{sec:berry}, starting with opposite electronic spin, the nuclei should feel equal and opposite Berry magnetic force. Therefore, the difference between spin-to-spin transmission rates should reveal some information about Berry curvature. 
Beyond raw transmission, we will also report the absolute spin polarization ($\Delta T$) and the relative spin polarization ($P$):
\begin{align}
   \Delta T = T_{\up\rarr \up} + T_{\down\rarr \up} - T_{\up\rarr \down} - T_{\down\rarr \down} \label{eq:polar1} \\
	P = \frac{\Delta T}{T_{overall}} = \frac{T_{\up\rarr \up} + T_{\down\rarr \up} - T_{\up\rarr \down} - T_{\down\rarr \down}}{T_{\up\rarr \up} + T_{\down\rarr \up} + T_{\up\rarr \down} + T_{\down\rarr \down}} \label{eq:polar2}
\end{align}
By investigating these three observables ($T$,$\Delta T$ and $P$) for a series of Hamiltonians (that break spatial symmetry in one way or another), we will necessarily learn about how Berry force affects spin-dependent chemical reaction rates.
%Obviously, we will need to study systems that break spatial inversion.

\subsection{Form of the Hamiltonian}

The Hamiltonians studied below will be characterized by three different criteria:

\subsubsection{Form of SOC}

The first criterion is the form of the SOC. Generally, a Hamiltonian with SOC has the form
\begin{align} \label{eq:soc}
    H_{SO} &= \alpha(\sigma_x \hat{L}_x + \sigma_y \hat{L}_y + \sigma_z \hat{L}_z) \nonumber \\ 
    &= \alpha\sum_{j,k}{{(L_x - iL_y)}_{jk}c_{j\up}^\dagger c_{k\down} + {(L_x + iL_y)}_{jk}c_{j\down}^\dagger c_{k\up} + {(L_z)}_{jk}(c_{j\up}^\dagger c_{k\up} - c_{j\down}^\dagger c_{k\down}) }
\end{align}
where $\alpha$ is SOC magnitude and $\hat{L}$'s are electronic angular momentum operators. Comparing Eq.~\eqref{eq:soc} with Eq.~\eqref{eq:hfull}, the two SOC terms in our model, $s_{\up\up}(\bm{R})$ and $s_{\up\down}(\bm{R})$, correspond to the $z$-component and $xy$-component of SOC, respectively. Here, we will assume that only one of $s_{\up\up}(\bm{R})$ and $s_{\up\down}(\bm{R})$ is nonzero, so that the Hamiltonian can be categorized into two cases:
\begin{enumerate}
	\item A two-state case: $s_{\up\up}(\bm{R}) \ne 0$, $s_{\up\down}(\bm{R}) = 0$, where the two opposite spins are decoupled. The effective Hamiltonian of each spin consists of two electronic states:
	\begin{align} \label{eq:twostateh}
    	H_{\up\up} = -\frac{\hbar^2}{2M}\nabla_{\bm{R}}^2 + \begin{bmatrix}
    	E_1(\bm{R}) & V_t(\bm{R}) \\ V_t^*(\bm{R}) & E_2(\bm{R})
    	\end{bmatrix},\quad H_{\down\down}=H_{\up\up}^*
	\end{align}
	
	\item A four-state case with $s_{\up\up}(\bm{R}) = 0$, $s_{\up\down}(\bm{R}) \ne 0$. Here, Berry curvature arises only from the complex nature of the spin-flipping SOC.
\end{enumerate}

\subsubsection{Geometry of Diabatic Potential Energy Surfaces}

The second criterion is the geometric form of the diabatic potential energy surfaces. Again we consider two cases: a bent geometry and a linear geometry. See Fig. \ref{figure:geometry}.
%We will define an activ%In both systems, the nuclei are confined in the active region $\bm{R} \in \mathbb{G}$, i.e. $H=+\infty$ when $\bm{R} \notin \mathbb{G}$.

The parameters for the bent system are
\begin{flalign} \label{eq:param1_1}
    %&\mathbb{G} = \{(R, \theta) |  R_{min} < R < R_{max} \text{ and } 0 < \theta < \frac\pi 2 \} \\
    &E_1(R,\theta) = A(1+\tanh(\epsilon(\theta-\frac\pi 4))) \\
    &E_2(R,\theta) = A(1-\tanh(\epsilon(\theta-\frac\pi 4))) \label{eq:param1_2}
\end{flalign}
where $R=\sqrt{x^2+y^2}$ and $\theta=\atan{(y/x)}$ are the polar coordinates. We place a hard wall at $R=R_{min}$ or  $R= R_{max}$, i.e. $E_1 = E_2 = +\infty$ if $R<R_{min}$ or $R > R_{max}$. As far as the couplings are concerned, we set:
\begin{align} \label{eq:parambent1}
    V_t(R,\theta) = \begin{cases}
    C \exp(-\epsilon^2(\theta-\frac\pi 4)^2+iWu(x,y)) & \text{ Two-state Case}\\
    C \exp(-\epsilon^2(\theta-\frac\pi 4)^2) & \text{ Four-state Case} \\
\end{cases} 
\end{align}
\begin{align} \label{eq:parambent2}
    V_s(R,\theta) = \begin{cases}
    0 & \text{ Two-state Case}\\
    D \exp(-\epsilon^2(\theta-\frac\pi 4)^2+iWu(x,y)) & \text{ Four-state Case} \\
    \end{cases} 
\end{align}
The incoming and outgoing terminals of the system are located at $\theta=0$ and $\theta=\pi/2$, respectively. We choose $R_{min} = 4$, $R_{max} = 8$, $\epsilon = 5$ and $M = 1000$ (all in atomic units, same below) in our simulations. Typically $D \ll C$ and $W$ in the two-state case is smaller than $W$ in the four-state case.

The parameters for the linear system are
\begin{align}
    %&\mathbb{G} = \{(x, y) | -\frac{L_y}2 < y < \frac{L_y}2 \text{ and } -\frac{L_x}{2} < x < \frac{L_x}{2} \} \\
    &E_1(x,y) = A(1+\tanh(\epsilon y)) \\
    &E_2(x,y) = A(1-\tanh(\epsilon y))
\end{align}
We place a hard wall at $x=-\frac{L_x}{2}$ or $x= \frac{L_x}{2}$, i.e. $E_1 = E_2 = +\infty$ if $x<-\frac{L_x}{2}$ or $x > \frac{L_x}{2}$.

As far as the couplings are concerned, we set:
\begin{align} \label{eq:paramlinear1}
    V_t(x,y) = \begin{cases}
    C \exp(-\epsilon^2y^2+iWu(x,y)) & \text{ Two-state Case}\\
    C \exp(-\epsilon^2y^2) & \text{ Four-state Case} \\
    \end{cases} 
\end{align}
\begin{align} \label{eq:paramlinear2}
    V_s(x,y) = \begin{cases}
    0 & \text{ Two-state Case}\\
    D \exp(-\epsilon^2y^2+iWu(x,y)) & \text{ Four-state Case} \\
    \end{cases} 
\end{align}
The incoming and outgoing terminals of the system are located at $y=-\frac{L_y}{2}$ and $y=\frac{L_y}{2}$, respectively. We choose $L_x = 8$, $L_y = 4$, $\epsilon = 5$ and $M = 1000$ in our simulations.

\subsubsection{Geometry dependence of the SOC Phase}

Our third criterion is the phase changing direction $u(x,y)$ (see Eq.~\eqref{eq:parambent1}, \eqref{eq:parambent2}, \eqref{eq:paramlinear1} and \eqref{eq:paramlinear2}). We investigate two types of phase modulation functions $u$:
\begin{enumerate}
	\item Radial dependence: $u(x,y) = \sqrt{x^2 + y^2}$
	\item Linear dependence: $u(x,y) = x\cos{\phi} + y\sin{\phi}$. Here $\phi$ is a parameter specifying the coupling direction.
\end{enumerate}

\subsection{Berry Curvature}
Recall that, for a system in 2D, the Berry curvature can be quantified by an effective magnetic field (see Eq.~\eqref{eq:bfield}). Here, we will give analytic expressions for that magnetic field along the adiabatic ground state.\cite{footnote1}

\subsubsection{Two-State Case}

In the two-state case, the spin is a well-defined quantity everywhere, so we can use spin to label the adiabatic ground surfaces. As discussed before, the equivalent magnetic field for up and down spin surfaces are equal and opposite.

For the case of a bent geometry, the equivalent magnetic field for adiabatic state $\ket{0\up}$ is given by
\begin{align} \label{eq:btwostatebent}
    \bm{B}_{0\up}(R,\theta) = -\dfrac{A \abs{V_t}^2}{{(\Delta^2 + \abs{V_t}^2)}^{3/2}} \dfrac{W\epsilon}{2R} (\sech^2{\tilde{\theta}} + 2\tilde{\theta}\tanh{\tilde{\theta}}) \gamma(\theta)\bm{e}_z
\end{align}
Eq.~\eqref{eq:btwostatebent} is proved in Appendix \ref{sec:appendixbfield}. For the case of linear geometry, the equivalent magnetic field for $\ket{0\up}$ is
\begin{align} \label{eq:btwostatestraight}
    \bm{B}_{0\up}(x,y) = -\dfrac{A \abs{V_t}^2}{{(\Delta^2 + \abs{V_t}^2)}^{3/2}} \dfrac{W\epsilon}{2} (\sech^2{\tilde{y}} + 2\tilde{y}\tanh{\tilde{y}}) \gamma'(x, y)\bm{e}_z
\end{align}
In Eqs.~\eqref{eq:btwostatebent} and \eqref{eq:btwostatestraight}, $\Delta = \frac{(E_2 - E_1)}{2}$, $\tilde{\theta} = \epsilon(\theta - \frac{\pi}{4})$, $\tilde{y} = \epsilon y$ and $\gamma(\theta),\gamma'(\theta)$ are unitless factors depending on $u(x,y)$:
\begin{align} \label{eq:gamma}
    \gamma(\theta) = \begin{cases}
        1 & u(x,y) = \sqrt{x^2+y^2} \\
        \cos{(\theta - \phi)} & u(x,y) = x\cos{\phi} + y\sin{\phi}
    \end{cases} \\
    \gamma'(x,y) = \begin{cases}
        \dfrac x {\sqrt{x^2 + y^2}} & u(x,y) = \sqrt{x^2+y^2} \\
        \cos{\phi} & u(x,y) = x\cos{\phi} + y\sin{\phi}
    \end{cases}
\end{align}

For the bent geometry, since the diabatic crossing point is located at $\theta = \frac{\pi}{4}$, where the diabatic coupling $V_t$ also reaches its maximum, it is worthwhile to report $\bm{B}_{0\up}$ at $\theta=\frac{\pi}{4}$ as a representative value of the Berry curvature strength:
\begin{align} \label{eq:bapprox}
    \bm{B}_{0\up}(R,\dfrac\pi 4) = -\dfrac{AW\epsilon}{2CR}\gamma(\dfrac\pi 4)\bm{e}_z
\end{align}

In all expressions above, the equivalent magnetic field for the $\ket{0\down}$ surface is $\bm{B}_{0\down}(x,y) = -\bm{B}_{0\up}(x,y)$.

\subsubsection{Four-state Case}

One can perform the same calculation as above for the four-state case. However, since the adiabatic ground state is doubly degenerate, and spin is not a good quantum number here, the definition of the two adiabatic ground states is left as a matter of choice. Here we choose our ground adiabatic states as (in the diabatic basis $\ket{1\up},\ket{1\down},\ket{2\up},\ket{2\down}$):
\begin{align}
    \ket{0\alpha} = \eta\Big[V_t, -V_s, \frac{\sqrt{{|V_t|}^2+{|V_s|}^2}}{\Delta-\sqrt{\Delta^2+\abs{V_t}^2+\abs{V_s}^2}}, 0\Big]^T \nonumber \\
    \ket{0\beta} = \eta\Big[V_s, V_t, 0, \frac{\sqrt{{|V_t|}^2+{|V_s|}^2}}{\Delta-\sqrt{\Delta^2+\abs{V_t}^2+\abs{V_s}^2}}\Big]^T
\end{align}
where $\eta$ is the normalization constant. In our simulations, we set $V_t \gg V_s$, so when $\theta \ll \frac{\pi}{4}$, $\ket{0\alpha} \approx \ket{1\up}$ and $\ket{0\beta} \approx \ket{1\down}$ and when $\theta \gg \frac{\pi}{4}$, $\ket{0\alpha} \approx \ket{2\up}$ and $\ket{0\beta} \approx \ket{2\down}$. Therefore, in the terminals, the $\alpha$ and $\beta$ states can be approximated as having up spin and down spin, respectively.

For the bent geometry, $\bm{B}_{0\alpha}$ is given by
\begin{align} \label{eq:bfourstatebent}
    \bm{B}_{0\alpha}(R,\theta) =  \dfrac{A \abs{V_s}^2}{{(\Delta^2 + \abs{V_t}^2 + \abs{V_s}^2)}^{3/2}} \dfrac{W\epsilon}{2R} (\sech^2{\tilde{\theta}} + 2\tilde{\theta}\tanh{\tilde{\theta}}) \gamma(\theta) \bm{e}_z
    \text{ ,}
\end{align}
and when $\theta=\frac{\pi}{4}$,
\begin{align} \label{eq:bapprox2}
    \bm{B}_{0\alpha}(R,\dfrac\pi 4) = \dfrac{AW\epsilon}{2R} \dfrac{D^2}{{(C^2+D^2)}^{3/2}}\gamma(\dfrac\pi 4)\bm{e}_z \text{ .}
\end{align}
For the linear geometry, $B_{0\alpha}$ is given by
\begin{align}
    \bm{B}_{0\alpha}(x,y) =  \dfrac{A \abs{V_s}^2}{{({\Delta E}^2 + \abs{V_t}^2 + \abs{V_s}^2)}^{3/2}} \dfrac{W\epsilon}{2} (\sech^2{\tilde{y}} + 2\tilde{y}\tanh{\tilde{y}}) \gamma'(x, y) \bm{e}_z
\end{align}

In all expressions above, $\bm{B}_{0\beta}(x,y) = -\bm{B}_{0\alpha}(x,y)$.

\section{Computational Results} \label{sec:results}

\subsection{Methods for Calculating Transmission Probabilities}

\begin{figure}[H]
	\begin{center}
		\includegraphics[width=0.6\columnwidth]{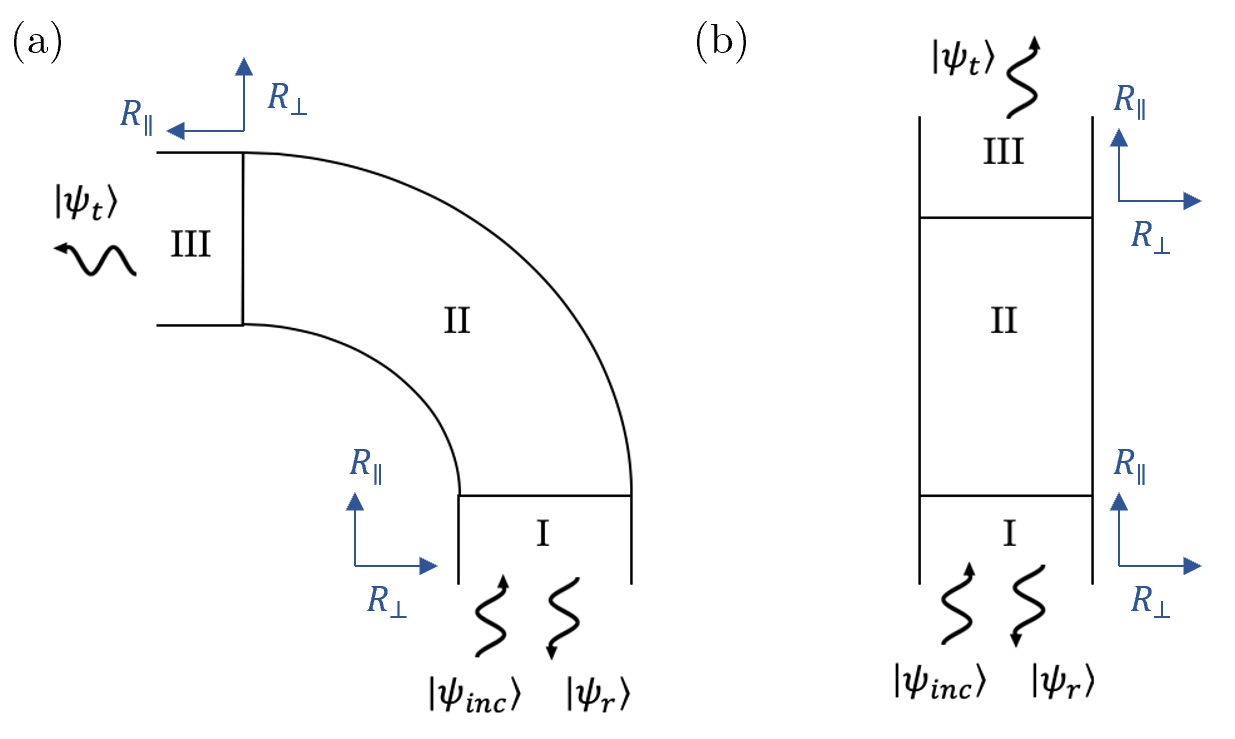}
	\end{center}
	\caption{Region setup for scattering calculation in (a) Bent geometry; (b) Linear geometry. The definitions of longitudinal coordinate ($R_\parallel$) and transverse coordinate ($R_\perp$) are marked in the figure. Note for the linear geometry, there is one unique $R_\parallel$ and one unique $R_\perp$.}
	\label{figure:geometry}
\end{figure}

To calculate transmission probabilities, we have run two-dimensional quantum scattering calculation, where we solve the Schrodinger Equation: \cite{Subotnik2008}
\begin{align} \label{eq:scattercore}
(H - E)\ket{\Psi} = 0
\end{align}
Here, $H$ is the model Hamiltonian, $E$ is the incoming energy and $\ket{\Psi}$ is the overall wave function. $\ket{\Psi}$ is defined in three regions:
\begin{align}
    \ket{\Psi(x,y)} =
    \begin{cases}
        \ket{\psi_{inc}(x,y)} + \ket{\psi_r(x,y)} & \text{ Region I} \\
        \ket{\psi_s(x,y)} & \text{ Region II} \\
        \ket{\psi_t(x,y)} & \text{ Region III} \\
    \end{cases}
\end{align}
Region II is the region of interest, and regions I/III are located (respectively) on the incoming/outgoing sides of region II. The region definition for each geometry is shown in Fig. \ref{figure:geometry}.

The four parts of the wave function are: an incoming wave $\ket{\psi_{inc}}$, a reflecting wave $\psi_r$, a transmitting wave $\ket{\psi_t}$ and the interior wave function $\psi_s$. Their specific expressions are defined as
\begin{align}
    \ket{\psi_{inc}(x,y)} =& e^{ik_{n_{inc} \mu_{inc}}^{\text{I}}y}\ket{n_{inc}}\otimes\ket{y}\otimes\ket{u_{n_{inc} \mu_{inc}}^{\text{I}}(x)} \\
    \ket{\psi_{r}(x,y)} =& \sum_{n,\mu}{r_{n\mu} e^{-ik_{n\mu}^{\text{I}}y}\ket{n}\otimes\ket{y}\otimes\ket{u_{n\mu}^{\text{I}}(x)}} \\
    \ket{\psi_{t}(R_\parallel,R_\perp)} =& \sum_{n,\mu}{t_{n\mu} e^{ik_{n\mu}^{\text{III}}R_{\parallel}}\ket{n}\otimes\ket{R_\parallel}\otimes\ket{u_{n \mu}^{\text{III}}(R_\perp)}} 
\end{align}
Here, $\ket{n}$ is the electronic state, $r_{n \mu}$ and $t_{n \mu}$ are the reflection and transmission coefficients in channel $n\mu$, $R_\parallel$ and $R_\perp$ is the longitudinal and transversal coordinates in region III. For the bent geometry $R_\parallel = -x$ and $R_\perp = y$; for the linear geometry, $R_\parallel = y$ and $R_\perp = x$. $\ket{u^{\text{I}}_{n \mu}}$ and $\ket{u^{\text{III}}_{n \mu}}$ are the $\mu^{\text{th}}$ transverse eigenstates associated with electronic state $n$ in regions I and III, which satisfies
\begin{align} \label{eq:eigcalc}
    (-\frac{\hbar^2}{2M} \frac{\partial^2}{\partial x^2} + V_{n}^{\text{I}})\ket{u_{n \mu}^r(x)} = E_{n \mu}^{\text{I}} \ket{u_{n \mu}^{\text{I}}(x)} \\
    (-\frac{\hbar^2}{2M} \frac{\partial^2}{\partial R^2_\perp} + V_{n}^{\text{III}})\ket{u_{n \mu}^t(R_\perp)} = E_{n \mu}^{\text{III}} \ket{u_{n \mu}^{\text{III}}(R_\perp)}       
\end{align}
Here $V_{n}^{\text{I}}$ and $V_{n}^{\text{III}}$ are the diabatic energies of electronic state $n$ at regions I and III. In regions I and III, the wave numbers associated with energies $E_0$, nuclear transverse state $\mu$ and electronic state $n$ are $k_{n \mu}^{\text{I}} = \sqrt{2M(E_0 - E_{n \mu}^{\text{I}})}$ and $k_{n \mu}^{\text{III}} = \sqrt{2M(E_0 - E_{n \mu}^{\text{III}})}$, respectively.

As discussed above, we fix the incoming transverse state to be the ground state (i.e. $\mu_{inc} = 0$) in all calculations. To calculate the spin-dependent transmission rates, we choose $n_{inc}$ to be either adiabatic state $\ket{0\up}$ or $\ket{0\down}$.

\subsection{Simulation Results of the Two-state System}

\begin{figure}[H]
	\includegraphics[width=\columnwidth]{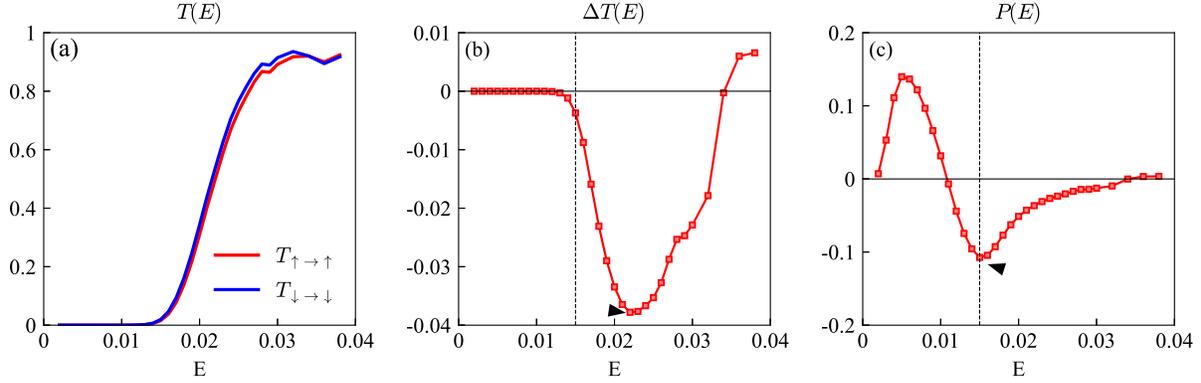}
	\caption{Two-state system simulation results for a bent geometry. (a) Spin-specific transmission rate $T(E)$, (b) transmission rate difference $\Delta T(E)$ and (c) polarization $P(E)$ as functions of energy; see Eqs.~\eqref{eq:polar1},\eqref{eq:polar2}. The energy barrier is shown by the vertical dashed line. The polarization peaks are marked with black triangles in (b) and (c). The parameters are $A=0.02$, $C=0.005$, $W=0.2$ and $u(x,y)=\sqrt{x^2+y^2}$. See Eqs.~\eqref{eq:param1_1},\eqref{eq:param1_2},\eqref{eq:parambent1}. Note that, while spin polarization is observed, the details of which spin is favored at which energy is rather complicated.}
	\label{figure:twostate}
\end{figure}

We begin our investigation by plotting the transmission and spin polarization for the two-state system ($V_t\ne 0$ and $V_s = 0$) with bent geometry. As shown in Fig. \ref{figure:twostate}a, we calculate the transmission rates $T_{\up\rarr\up}$ and $T_{\down\rarr\down}$ with model parameters $A=0.02$, $C=0.005$, $W=0.2$. We set $u(x,y)=\sqrt{x^2+y^2}$ and the overall incoming energy $0.001\le E\le 0.038$. As would be expected for any scattering process through or over an energetic barrier, the transmission is close to zero for both spins when the incoming energy is less than the barrier height (which is given by $E_{barrier}=A-C=0.015$ here); vice versa, when energy is over the barrier height, the transmission increases dramatically.

In Fig. \ref{figure:twostate}b, we plot the absolute spin polarization $\Delta T$, as calculated by Eq.~\eqref{eq:polar1} at each incoming energy. For most incoming energies ($0.010 < E < 0.034$), the down spin is significantly preferred. The absolute polarization reaches its maximum magnitude ($0.038$) when the incoming energy is slightly above the barrier ($E\approx 0.022$ here), and decreases when the incoming energy gets even higher. Not intuitively, when the incoming energy approaches the threshold for populating the first excited state ($E > 0.034$), the up spin is preferred.

Lastly, in Fig. \ref{figure:twostate}c, we plot the relative spin polarization $P$ as calculated by Eq.~\eqref{eq:polar2}. Around $E= 0.02$, $P$ follows the same pattern as $\Delta T$ in so far as the down spin is preferred, though the peak minimum is shifted to a lower energy (around $E=0.015$, as marked by the triangles). For very low energies, we find that the transmission of the up spin is actually preferred; however, any preference at such a low energy is likely not very meaningful as the total transmission is effectively negligible.

\subsubsection{Polarization Dependence on the Phase Gradient Magnitude, Energy Gap and Coupling Strength} \label{sec:twostateparam}

\begin{figure}[H]
	\begin{center}
		\includegraphics[width=\columnwidth]{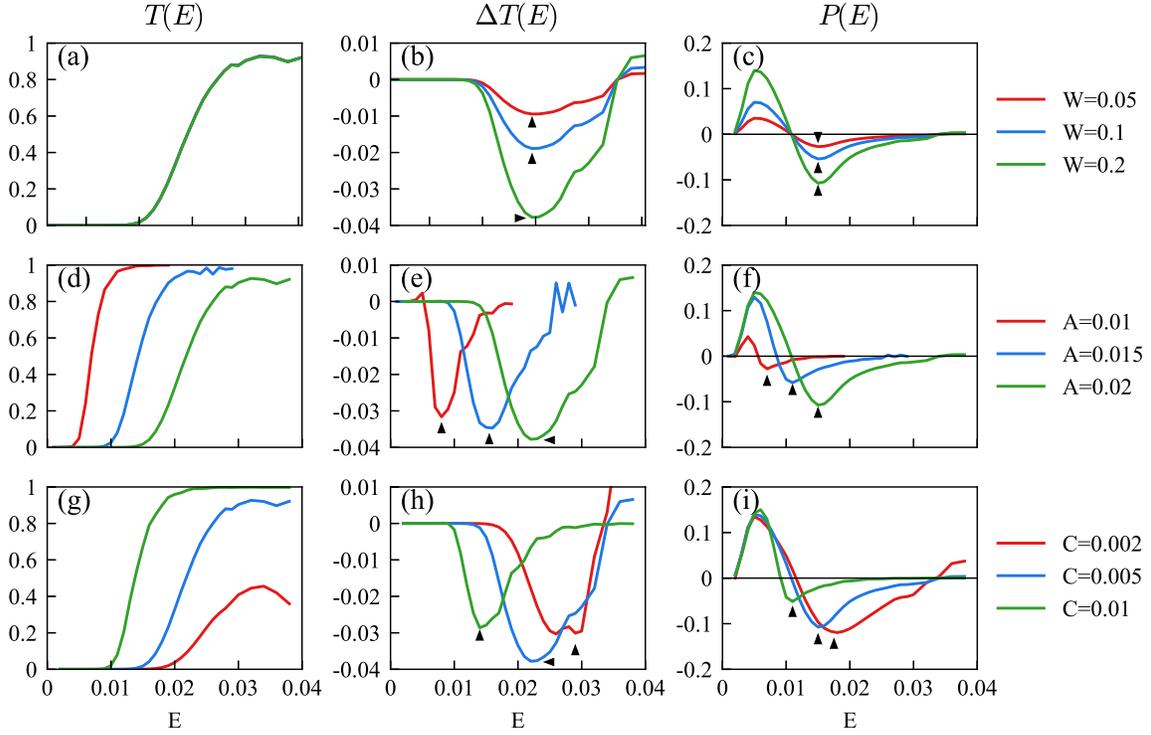}
	\end{center}
	\caption{The total transmission $T(E)$ (a,d,g), absolute polarization $\Delta T(E)$ (b,e,h) and relative polarization $P(E)$ (c,f,i) with different phase gradient magnitude $W$, half energy gap $A$ and diabatic coupling strength $C$ for a two-state system with a bent geometry. The polarization peaks (which will be plotted in Fig. \ref{figure:twostatecmp2} below) are marked with black triangles (b,c,e,f,h,i). The default parameters are $A=0.02$, $C=0.005$, $W=0.2$ and $u(x,y) = \sqrt{x^2+y^2}$. See Eqs.~\eqref{eq:param1_1},\eqref{eq:param1_2},\eqref{eq:parambent1}. Increasing $W$ does not change transmission but does amplify polarization (a-c); Increasing $A$ shifts both transmission and polarization (d-f); Increasing $C$ shifts transmission and polarization, but also reduces transmission (g-i).}
	\label{figure:twostatecmp}
\end{figure}

Having identified spin polarization for one Hamiltonian, we will now turn our attention to the more important question: how does spin polarization change for different Hamiltonians?
Specifically, we will focus on three parameters: the phase gradient magnitude $W$, the half energy gap $A$ and the coupling strength $C$. We will vary one parameter at a time, and fix all other parameters as their default values: $W=0.2$, $A=0.02$ and $C=0.005$. Results are plotted in Fig. \ref{figure:twostatecmp}.
As one might expect, we find that changing the phase gradient $W$ has no influence on transmission (Fig. \ref{figure:twostatecmp}a), but this change does affect the polarization (both absolute and relative) dramatically (Fig. \ref{figure:twostatecmp}b,c). 
By contrast, increasing the half energy gap $A$ both shifts the transmission and leads to a broadened and deepen polarization peak (Fig. \ref{figure:twostatecmp}d-f); The former result is intuitive because changing $A$ leads to a change in the energy barrier; the latter result is more interesting and will be discussed below.
Finally, changing the coupling strength $C$ also affects both transmission and polarization (Fig. \ref{figure:twostatecmp}g-i). As the diabatic coupling strength increases, the transmission increases and shifts to low energy, because, with a strong coupling strength, the energy barrier of the adiabatic ground state is lower and the system becomes more adiabatic. However, more interestingly, as $C$ increases, the relative polarization becomes smaller, as will also be discussed below in detail.

\begin{figure}[H]
	\begin{center}
		\includegraphics[width=\columnwidth]{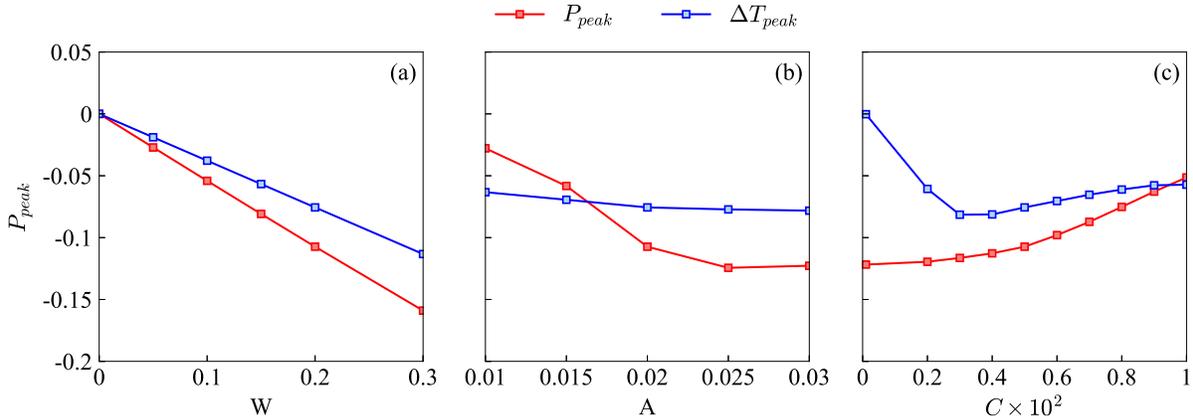}
	\end{center}
	\caption{Peak relative polarization $P_{peak}$ and peak absolute polarization $\Delta T_{peak}$ (see black arrows in Fig. \ref{figure:twostate}c and Fig. \ref{figure:twostate}b) as functions of $W$,$A$,$C$ for the two-state system with bent geometry. The default parameters are $A=0.02$, $C=0.005$, $W=0.2$ and $u(x,y) = \sqrt{x^2+y^2}$. The magnitude of $P_{peak}$ increases with $W$ and $A$ but decreases with $C$, which agrees with the strength of effective magnetic field (Eq.~\eqref{eq:bapprox}) qualitatively. Thus, the relative peak polarization goes up in the nonadiabatic limit. Vice versa, the peak absolute polarization has a maximum around $C=0.003$.}
	\label{figure:twostatecmp2}
\end{figure}

To better analyze the data in Fig. \ref{figure:twostatecmp}, in Fig. \ref{figure:twostatecmp2} we plot (in red) the peak value for the relative polarization $P_{peak}$ (i.e. which is defined as the extreme value of the relative polarization $P(E)$ when $E > 0.9E_{barrier}$ [see the black triangles in Fig. \ref{figure:twostatecmp}]) as a function of the three parameters $W,C,A$. We also plot (in blue) the peak in absolute polarization $\Delta T(E)$.

We begin with $P(E)$. As would be expected qualitatively from the analytic expression for the effective magnetic field in Eq.~\eqref{eq:bapprox}, we find the following:
\begin{itemize}
	\item The peak polarization $P_{peak}$ has linear dependence on $W$ (Fig. \ref{figure:twostatecmp2}a).
	\item The peak polarization becomes smaller as the diabatic coupling strength $C$ increases (Fig. \ref{figure:twostatecmp2}b) and one enters the nonadiabatic limit.
	\item The peak polarization $P_{peak}$ increases with half energy gap $A$  (Fig. \ref{figure:twostatecmp2}c).
\end{itemize}
Note that interpreting the last results above is difficult: $A$ is not as simple as just changing the adiabaticity of the rate process because $A$ modulates both the asymptotic energy gap as well as the gradients of the diabatic curves at the crossing point. Overall, the spin-polarization appears to track very well with the form of the effective magnetic field at the crossing point (Eq.~\eqref{eq:bapprox}), and our preliminary conclusion is that relative spin-polarization should be largest in the nonadiabatic limit. 

Next, we turn to $\Delta T(E)$. As far as dependence on $W$ and $A$, the peak in $\Delta T(E)$ follows the peak in $P(E)$ qualitatively. However, realizing that the total transmission must go to zero in the extreme nonadiabatic limit (when $C\rightarrow 0)$, it makes sense that we find that the peak in absolute spin polarization (i.e. the maximum $\Delta T(E)$) has a maximum for an intermediate value of $C$.
Thus, for the experimentalist interested in measuring or producing large spin polarization, there is a sweet spot for the Hamiltonian.

\subsubsection{Polarization Dependence on the Phase Gradient Direction and Geometry of Energy Surface}

Having investigated how spin polarization depends on the {\em magnitude} of the phase gradient of the diabatic coupling ($W$), let us next turn our attention to the {\em direction} of the phase gradient. We will work with both the bent and linear geometries.

\begin{figure}[H]
	\includegraphics[width=\columnwidth]{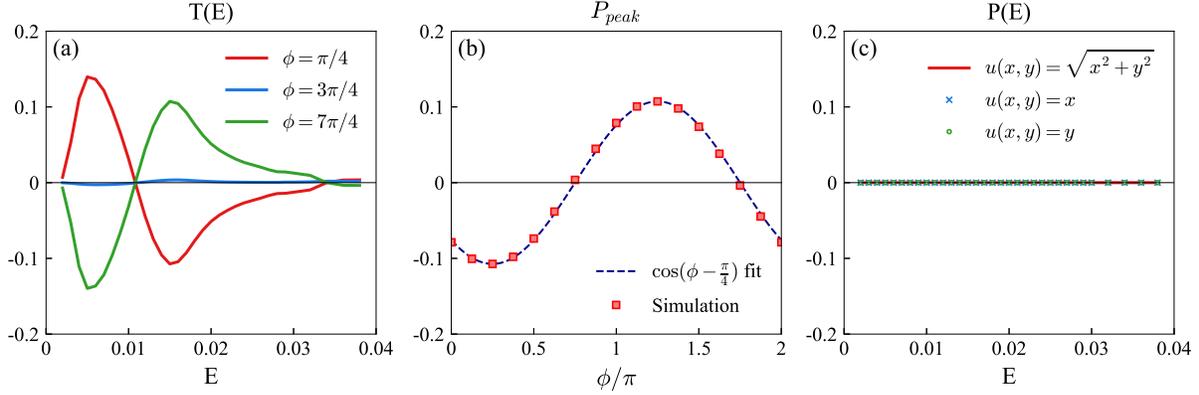}
	\caption{(a) Energy-polarization curve for the bent geometry for a few different phase directions $\phi$. 
    (b) $P_{peak}$ as function of $\phi$ for the bent geometry. Note that for the bent systems, the overall spin polarization can be mapped directly onto the Berry force at the crossing point (see Eqs.~\eqref{eq:gamma} and \eqref{eq:bapprox}), denoted ``$\cos{(\phi-\frac{\pi}{4})}$ fit'' here. 
    (c) Energy-polarization curve for the linear system for a few different phase dependence functions $u(x,y)$. Note here that, even though the Berry force has the same form as in Fig. \ref{figure:twostatespecial}a, there is no spin polarization (which highlights the fact that the observed spin polarization depends on the both the magnitude of the Berry force and the shape of the potential energy surfaces). The default parameters for all systems are $A=0.02$, $C=0.005$ and $W=0.2$.}
	\label{figure:twostatespecial}
\end{figure}

First, for the case of a bent geometry, we will set perform simulations for a family of possible phases $u(x,y) = x \cos{\phi} + y \sin{\phi}$ where $\phi \in [0, 2\pi]$. 
In Fig. \ref{figure:twostatespecial}a, we show some representative results for $P(E)$ with different angles $\phi$, and in Fig. \ref{figure:twostatespecial}b, we plot the dependence of $P_{peak}$ on $\phi$.
Overall, we find that the polarization  $P(E)$  is largest when $\phi=\frac{\pi}{4}$ or $\frac{7\pi}{4}$, i.e. where the phase gradient is perpendicular to the momentum of nuclei. Vice versa, when $\phi=\frac{3\pi}{4}$ or $\frac{11\pi}{4}$, the phase gradient is parallel to the momentum, and the polarization is almost zero.
As above, our result can be explained by considering the analytic expression for the effective magnetic field at the crossing point as shown in Eqs.~\eqref{eq:gamma} and \eqref{eq:bapprox}, the effective magnetic field at a diabatic crossing point is proportional to $\cos{(\phi-\frac{\pi}{4})}$, and both the polarization and the magnetic field are maximized when $\phi=\frac{\pi}{4}$ or $\frac{7\pi}{4}$.
Note that, in Fig. \ref{figure:twostatespecial}b,  we have fit the  computed $P_{peak}-\phi$ curve by $\cos{(\phi-\frac{\pi}{4})}$, and we find that the relative polarization agrees quantitatively with the strength of the effective magnetic field at the crossing point (Eq.~\eqref{eq:bapprox}).

Second, we turn to the case of a linear geometry, where $x$ is the transverse direction and $y$ is the direction of incoming momentum. Again, we perform simulations for the same family of possible phases.  Note that, in  Fig. \ref{figure:twostatespecial}c (and unlike Figs. \ref{figure:twostatespecial}a,b), we now observe no polarization for any $u(x,y)$ -- even though there is a nonzero Berry force (see Eq.~\eqref{eq:btwostatestraight}).
To rationalize this finding, note that the diabatic potential energies in this system have a plane of reflection symmetry that is parallel to the direction of incoming  and outgoing momentum.  In other words, even though incoming nuclei moving in the $y$ direction will feel opposite Berry magnetic forces in the $x$ direction (depending on their spin [up or down]) and therefore bend either to left ($-x$) or right ($+x$), their total transmission in the $y$ direction will not depend on the spin.\cite{footnote2}
Overall, Fig. \ref{figure:twostatespecial}c must remind us that, even though the spin polarization can be mapped to the effective magnetic field in Eq.~\eqref{eq:bapprox}, systems with same Berry curvature magnitude can yield completely different observable spin current behavior; after all, quantum wave packet motion depends just as much (and usually far more) on the adiabatic potential energy surfaces than it depends on overall Berry force.

This completes our discussion of the two state results.

\subsection{Simulation Results of the Four-state System}

\begin{figure}[H]
	\includegraphics[width=\columnwidth]{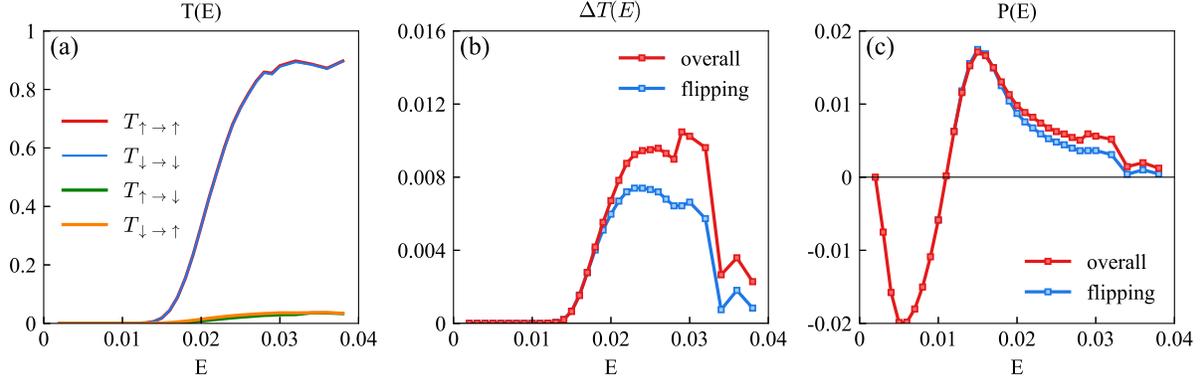}
	\caption{Four-state system simulation results. (a) Spin-specific transmission rate $T(E)$, (b) transmission rate difference $\Delta T(E)$ and (c) polarization $P(E)$ as functions of energy. In this figure, the `overall' $\Delta T(E)$ and $P(E)$ lines refer to Eq.~\eqref{eq:polar1} and Eq.~\eqref{eq:polar2}, respectively; the `flipping' $\Delta T(E)$ and $P(E)$ lines refer to the contribution of spin-flipping transmission to the polarization $T_{\down\rarr\up} - T_{\up\rarr\down}$, $(T_{\down\rarr\up} - T_{\up\rarr\down})/T_{overall}$, respectively. The parameters are $A=0.02$, $C=0.005$, $D=0.001$, $W=0.2$ and $u(x,y) = \sqrt{x^2+y^2}$. See Eqs.~\eqref{eq:param1_1},\eqref{eq:param1_2},\eqref{eq:parambent1},\eqref{eq:parambent2}. Note that, although the total transmission is dominated by spin-preserving processes, the spin polarization arises mostly from spin-flipping processes.}
	\label{figure:fourstate}
\end{figure}

Let us next turn our attention to the four-state systems with bent geometry. As above, we perform scattering calculations with incoming nuclear energy $0.001\le E\le 0.038$; the system parameters are $A=0.02$, $C=0.005$, $D=0.001$, $W=0.2$ (see Eq.~\eqref{eq:param1_1},\eqref{eq:param1_2},\eqref{eq:parambent1}, \eqref{eq:parambent2}).  Note that we now have two diabatic coupling parameters, $C$ and $D$. We set $u(x,y) = \sqrt{x^2+y^2}$. The data is plotted in Fig. \ref{figure:fourstate}. 
For the four-state system (and unlike the two-state system), one allows for spin-flipping transmissions, and so for a complete story, we must separately analyze all four cases for transmission rates between individual initial and final spins. See Fig. \ref{figure:fourstate}a.
We find that the energy-transmission curve for the four-state system has a pattern which is similar to the curve for the two-state system. The major contribution for transmission comes from spin-preserving transmission -- since the spin-preserving coupling $V_t$ is much larger than spin-flipping coupling $V_s$ here ($C \gg D$).

In Fig. \ref{figure:fourstate}b,c, we calculate the absolute and relative polarizations for the transmission data and find a preference for the transmission of spin up.
Here, we analyze in detail the contribution from spin-flipping transmission $\Delta T_{flip} = T_{\down\rarr\up} - T_{\up\rarr\down}$ and $P_{flip} = \frac{\Delta T_{flip}}{T_{overall}}$.
Even though spin-preserving transmission dominates spin-flipping transmission, we find that the major contribution to spin polarization comes from spin-flipping transmission. This is perhaps not so surprising since, although $C \gg D$, only $D$ multiplies a diabatic coupling with complex character.
%Lastly, as a point of distinction note that, contrary to the two-state case, one must be careful
%in interpreting the data in 
% because the Berry phase effect now mainly affect spin-flip transmissions, the up spin now gets favored, , .

\subsubsection{Polarization Dependence on the Phase Gradient Magnitude, Energy Gap and Coupling Strength}

\begin{figure}[H]
	\includegraphics[width=\columnwidth]{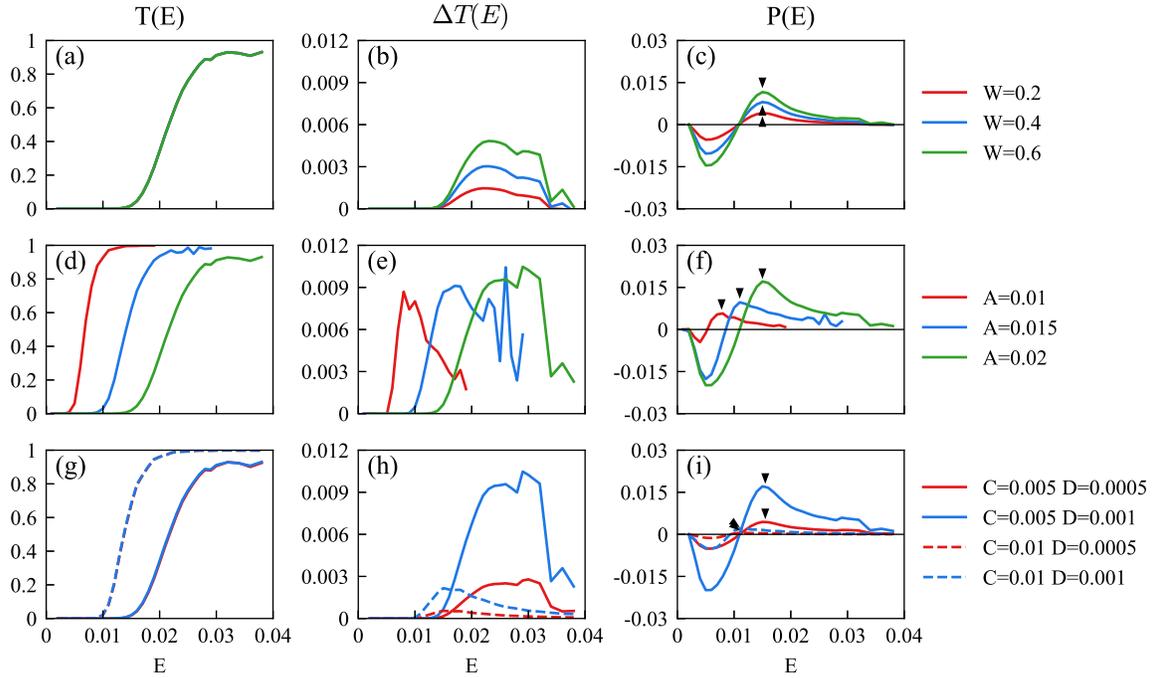}
	\caption{The total transmission $T(E)$ (a,d,g), absolute polarization $\Delta T(E)$ (b,e,h) and relative polarization $P(E)$ (c,f,i) with different $W$,$A$,$C$ and $D$ for the four-state systems with bent geometry. The polarization peaks (which will be plotted in Fig. \ref{figure:fourstatecmp2} below) are marked with black triangles (c,f,i). The default parameters are $A=0.02$, $C=0.005$, $D=0.001$, $W=1$ and $u(x,y) = \sqrt{x^2+y^2}$. See Eqs.~\eqref{eq:param1_1},\eqref{eq:param1_2},\eqref{eq:parambent1},\eqref{eq:parambent2}. The polarization for the four-state system is more sensitive to $C$ (relative to the two-state case) and increases with $D$.}
	\label{figure:fourstatecmp}
\end{figure}

At this point, again we turn to the parameter dependence of spin polarization for the four-state system with bent geometry. In addition to the phase changing rate $W$, the half energy gap $A$, the spin-preserve coupling strength $C$, there is another parameter of interest: the spin-flipping coupling strength $D$. As for the two-state case, we will vary one or two parameters at a time, and fix all other parameters as their default value: $A=0.02$, $C=0.005$, $D=0.001$, $W=1$. The results are plotted in Fig. \ref{figure:fourstatecmp}.
Similar to the two-state case, increasing $W$ does not change transmission, but dramatically increases polarization (Fig. \ref{figure:fourstatecmp}a-c); increasing $A$ shifts both transmission and polarization to higher energies (Fig. \ref{figure:fourstatecmp}d-f). 
The dependence of spin polarization on coupling strengths $C$,$D$ is more complicated. Two points are worth noting:
\begin{itemize}
	\item The transmission increases with $C$ and spin polarization decreases with $C$, similar to two-state system (Fig. \ref{figure:fourstatecmp}g-i). However, quantitatively speaking, the spin polarization (both absolute and relative) is more sensitive to $C$ in the four-state case relative to the two-state case. For instance, by doubling the value of $C$ from $0.005$ to $0.01$, the absolute polarization is reduced by a full $4/5$ of it original value. (whereas for the two-state case, one would expect a decrease of only $1/4$). This sensitivity must arise somehow from the competition between spin-preserving and spin-flipping transmissions, since the spin-flipping transmission is the major source of spin polarization. 
	\item As far as the spin-flipping coupling $D$ is concerned, increasing $D$ does not strongly affect the total transmission, but the polarization increases dramatically.
\end{itemize}

\begin{figure}[H]
\begin{center}
	\includegraphics[width=0.6\columnwidth]{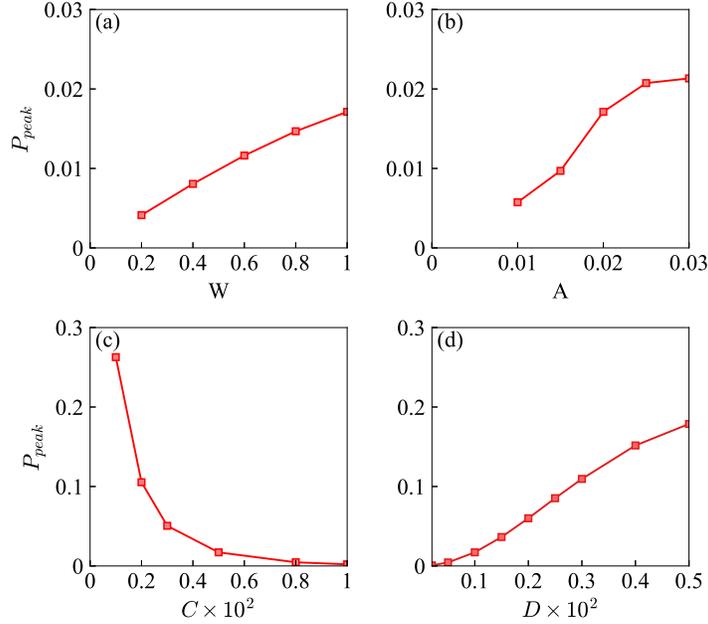}
	\caption{Relative peak polarization $P_{peak}$ as a function of the model parameters $W$,$A$,$C$,$D$ for the four-state system with a bent geometry. The default parameters are $A=0.02$, $C=0.005$, $D=0.001$, $W=1$ and $u(x,y) = \sqrt{x^2+y^2}$. Note that $P_{peak}$ increases with $W$,$A$,$D$ but decreases with $C$, which agrees with Eq.~\eqref{eq:bapprox2} qualitatively.}
	\label{figure:fourstatecmp2}
\end{center}
\end{figure}

Next, let us turn to the dependence of peak polarization $P_{peak}$ on the different parameters of the Hamiltonian. In Fig. \ref{figure:fourstatecmp2}, we plot $P_{peak}$ as a function of $W$,$A$,$C$,$D$. Here, $P_{peak}$ increases with $W$, $A$ and $D$ and decreases with $C$, which agrees qualitatively with the expression for the magnetic field at the crossing point (Eq.~\eqref{eq:bapprox2}).
It is worth mentioning that, according to Eq.~\eqref{eq:bapprox2}, the relative ratio between $C$ and $D$ is more important than the absolute value of each parameter as far as determining the amount of polarization. In other words, given only a small SOC, the spin polarization can still be very significant if the non-SOC coupling is small enough.

\section{Discussion} \label{sec:discussion}

From the data above, it is easy to conclude that Berry force can have a measurable effect on spin-polarized transmission (at least for a model system). Armed with this knowledge, there are quite a few outstanding items that we must highlight as unanswered questions -- questions that will hope to address over the next few years.

\subsection{Measuring Spin-Dependent Rate Constants in the Condensed Phase} \label{sec:discusscondensed}

The first outstanding question that we must address in the future is whether or not the spin polarization rates calculated above can be observed in the condensed phase (as opposed to the gas phase) with finite temperature and some degree of disorder or friction.
To begin our discussion, consider a system with nuclear and spin degrees of freedom where spin-flipping is not allowed and where one operates in the nearly adiabatic limit. In the classical limit, the leading correction to the dynamics will be the introduction of the Berry force in Eq.~\eqref{eq:berryforce}.
However, for any classical system, the presence of a magnetic field does not affect the equilibrium state according to the Fokker-Planck equation.\cite{footnote3} And as far as equilibrium rates are concerned, for realistic experiments in the condensed phase, there will be a distribution of momenta for the reactants and it is possible that all Berry force effects will be washed away entirely.
After all, according to the transition state theory (TST) of chemical reactions, one assumes an equilibrated velocity distribution across the barrier that separates reactants from products and the presence of a magnetic field has no influence at all on the rate; in general, this assumption seems to match most experiments.\cite{nitzan2006chemical} Dynamical corrections to TST are well known and could in principle depend on a magnetic field; but in general, dynamical corrections (like Kramers' theory\cite{Pollak1989,Hanggi1990}) are small and, of course, any such correction would also need to keep the equilibrium distribution unchanged by the magnetic field. 
Lastly, in the limit of an overdamped reaction, momenta equilibrate very quickly and no magnetic field could possibly be detected. Thus, if we think of a spin-dependent reaction as occurring on one classical surface, there cannot be any magnetic field effect on the equilibrium rate constant.

Next, consider a slightly more quantum-mechanical context where we imagine a curve-crossing (as in the discussion section above).  
As we showed in Sec. \ref{sec:twostateparam} above, we expect the relative spin-polarization to be maximized in the nonadiabatic limit. And yet, a very simple argument can demonstrate that in the nonadiabatic limit, the Fermi's golden rule (FGR) rate constant will not lead to spin polarization.
After all, consider the reaction from state \#1 to state \#2 initially without spin. The FGR rate is
\begin{align}
    k_{1\rarr 2} = \frac{2\pi}{\hbar Z_1}\sum_{\mu,\mu'}{e^{-\beta E_{1\mu}}\abs{V_{1\mu 2\mu'}}^2 \delta (E_{\mu'}-E_\mu)}
\end{align}
Notice that, since the rate depends only on the absolute value of couplings $V_{1\mu 2\mu'}$, changing the phases of the diabatic coupling does not affect reaction rates; Fermi's golden rule does not depend on whether the coupling is $V_{1\mu 2\mu'}$ or $V_{1\mu 2\mu'}^*$. With this in mind, suppose one allows spin to enter the dynamics. By time reversibility, we know that we must find:
\begin{align}
    \mel{1,\mu\sigma}{V}{2,\mu'\sigma'} = \mel{1,\mu\tilde{\sigma}}{V^*}{2,\mu'\tilde{\sigma}'} = \mel{1\mu \tilde{\sigma}}{V}{2\mu'\tilde{\sigma}'}^*
\end{align}
Therefore, the relevant FGR rate constant is now:
\begin{align}
    k_{1 \sigma \rarr 2} &= \frac{2\pi}{\hbar Z_1}\sum_{\mu,\mu',\sigma'}{e^{-\beta E_{1,\mu}}\abs{V_{1\mu\sigma 2\mu'\sigma'}}^2 \delta (E_{\mu'}-E_\mu)} \nonumber \\
    &= \frac{2\pi}{\hbar Z_1}\sum_{\mu,\mu',\sigma'}{e^{-\beta E_{1,\mu}}\abs{V^*_{1\mu\sigma 2\mu'\sigma'}}^2 \delta (E_{\mu'}-E_\mu)} \nonumber \\
    &= \frac{2\pi}{\hbar Z_1}\sum_{\mu,\mu', \tilde{\sigma'}}{e^{-\beta E_{1,\mu}}\abs{V_{1\mu\tilde{\sigma} 2\mu'\tilde{\sigma'}}}^2 \delta (E_{\mu'}-E_\mu)} \nonumber \\
    &= k_{1 \tilde{\sigma}\rarr 2\tilde{\sigma'}} 
\end{align}
Therefore, $k_{1\up \rarr 2} = k_{1\down \rarr 2}$ so that, according to FGR, if one starts in equilibrium in terminal \#1 with one spin orientation (up or down), the total probability of reaching terminal \#2 cannot depend on that initial spin orientation.

Now admittedly, the FGR argument presented above need not be correct in general; one can certainly imagine that spin-polarization arises at a level of theory beyond 2nd order perturbation theory (i.e. the basis for FGR).
Nevertheless, even so, from the classical argument above alone, one could make a reasonably strong case that Berry phase and Berry force effects for the nuclei should not yield any meaningful spin-dependence in the condensed phase -- at least as far the equilibrium rate constants are considered only.
That being said, however, for a system out of equilibrium (say, with a temperature gradient or a current running through\cite{Segal2003,Galperin2007a,Solomon2010,Galperin2008,Craven2018}), the arguments above become entirely void and, no matter how we do the calculation, we have every reason to expect a Berry force effect even in the condensed phase. Indeed, there is already a substantial literature that has investigated Berry forces in the context of molecular conduction experiments.\cite{Dou2015a,Galperin2015,Bode2012}
While we are currently working on the theory of non-equilibrium spin-dependent rate constants, it is already clear in practice that modern experiments do exist showing that spin-polarization can be promoted by the presence of a voltage\cite{Xie2011,Kettner2015,Kiran2016,Kettner2018,Naaman2020} -- which brings us necessarily to the CISS effect.\cite{Naaman2012, Naaman2019}

\subsection {Implications for CISS Effect}

\begin{figure}[H]
	\begin{center}
		\includegraphics[width=0.25\columnwidth]{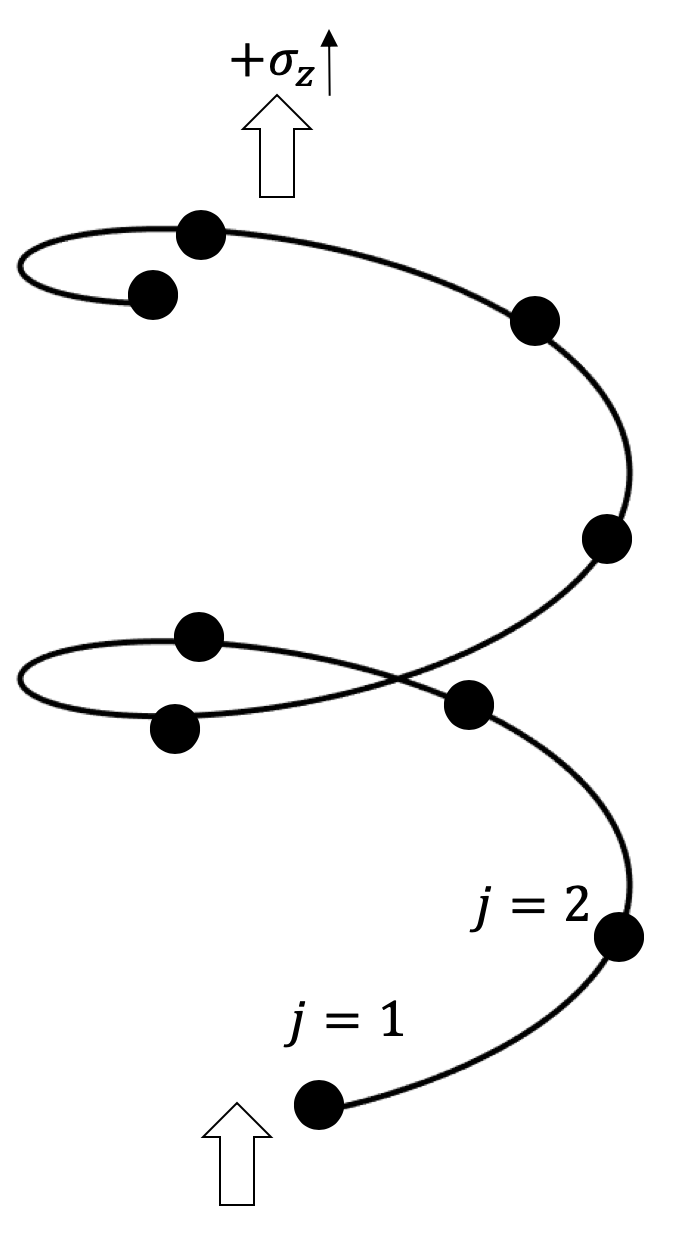}
	\end{center}
	\caption{A schematic model of the original CISS effect: an electron moving through a chiral potential has a probability of transmission that depends on its spin. The sites on the molecule are numbered.}
	\label{fig:helix}
\end{figure}

Although we will examine CISS effects in more detail in a later paper, a few brief words of review are perhaps appropriate regarding CISS (which represents the second outstanding question for the current research). As shown by Naaman and co-workers, the original CISS effect was demonstrated by the following scenario:
Imagine that one shines light on a gold (Au) surface coated with (chiral) dsDNA. If one measures the spin of the electrons ejected, one finds a spin preference for one spin orientation versus another (all relative the principle axis of the dsDNA).\cite{Gohler2011}
And over the years, it has been shown that this effect is not limited to dsDNA on gold: the effect occurs for various molecules or materials.\cite{Gohler2011,Xie2011,Kettner2015,Zwang2016,Eckshtain-Levi2016,Kiran2016,Abendroth2017,Aragones2017,Bloom2017,Suda2019,Inui2020,Naaman2020}
The basic molecular model of CISS effect is shown in Fig. \ref{fig:helix}: somehow or another (and the underlying physics remain debatable), an electron undergoing chiral transmission has a fundamental spin preference.

A multitude of theories have been proposed to explain the CISS effect \cite{Yeganeh2009,Medina2012,Guo2012,Guo2014,Gutierrez2012,Gutierrez2013,Rai2013,Gersten2013,Eremko2013,Kuzmin2014,Medina2015,Matityahu2016,Maslyuk2018,Naaman2019,Dalum2019,Zollner2020}. Indeed, the physics behind CISS have been and continue to be explored using approaches including (i) tight-binding atomistic simulations (where one assumes Rashba-like SOC) \cite{Guo2012,Guo2014,Gutierrez2012,Gutierrez2013,Rai2013} as well as (ii) solid state calculations, whereby one investigates transform in the context of band theory with SOC and a chiral potential.\cite{Eremko2013,Kuzmin2014,Medina2015,Matityahu2016,Naaman2019}
However, while all these theoretical frameworks successfully predict spin polarization in certain conditions, the fundamental physical force behind the CISS is not yet fully understood. Two essential items remain not fully explained:
\begin{enumerate}
	\item The magnitude of the observed spin polarization is not yet consistent with theory. For most atomistic calculations, a SOC much larger than a free carbon atom (several meV) is required for theory to match experiment. For some models,\cite{Medina2015,Naaman2019} one must also require that the energy of the incoming electron be comparable to the SOC gap; and yet, experimentally, CISS is observed for systems with thermal energy much higher than SOC energy.\cite{Gohler2011,Xie2011,Kettner2015,Kettner2018}
	\item For most theoretical models, calculations have not usually modeled incoherent effects present for charge transport. And yet, for molecules like DNA and protein, studies have shown that electron transfer can have a very large incoherent component.\cite{Zhang2014,Xiang2015,Kim2016,Beratan2019} More broadly, the overall effect of nuclear motion has not yet been fully explored for CISS.
\end{enumerate}

With this background in mind, the present paper offers another perspective on the CISS effect with regards to both of the questions above. 
In particular, here we have studied exactly how nuclear motion can become entangled with spin-dependent electronic dynamics, and we have shown that nuclear motion can indeed lead to quantitative and qualitative differences in spin-dependent transmission.
For instance, for one model system above, we find that
%The general electronic Hamiltonian of CISS reads
%\begin{align}
%   	H = \sum_{a,\sigma}{E_a c^\dagger_{a\sigma} c_{a\sigma}} + \sum_{a\ne b,\sigma}{t_{ab}c^\dagger_{a\sigma}c_{b\sigma}} + \sum_{a\ne b,\sigma,\sigma'}{s_{a\sigma b\sigma'}c^\dagger_{a\sigma} c_{b\sigma'}}
%    \label{eq:hciss}
%\end{align}
%where $a,b=1,\dots,N$. 
%The model Hamiltonian in Eq.~\eqref{eq:hfull} corresponds roughly to a two-site CISS model system with nuclear motion. If the nuclei and electron are decoupled, i.e. $E_a$, $t_{ab}$ and $s_{a\sigma b\sigma'}$ are constants in Eq.~\eqref{eq:hfull}, our model degenerates to a 4-by-4 Hamiltonian where apparently no spin polarization can happen. Therefore, our simulations indicate that at least in certain systems, nuclear motion is essential to the spin polarization.
polarization of \textasciitilde 10\% for some energies can be obtained with parameters that are not too exaggerated relative to the parameters used by other electronic-based models of CISS.\cite{Guo2012,Gutierrez2012} Thus, one must wonder what we will find if we begin to study CISS and properly include nuclear motion (as we do here).  
Interestingly, as discussed in Sec. \ref{sec:twoterminal}, we also might expect (in general) spin polarization to be most observable when we do not have equilibrated initial conditions in any one terminal,
e.g. what one might expect to see when running CISS experiments under a voltage (with current) or under illumination. Note that, according electron conduction studies, spin-polarization does increase dramatically far away from equilibrium (i.e. with high voltage).\cite{Xie2011,Kettner2015,Kiran2016,Kettner2018,Naaman2020}

Now, despite all of these enticing features, we must also admit that it is far too early to conclude whether or not (in any definitive way) nuclear motion directly connects to the actual CISS experiments.
After all, consider the na\"{i}vete of the simulations above:
$(i)$ we modeled unbound nuclear motion above (i.e. a nuclear scattering problem), which does not line up at all with Naaman's experiments;
$(ii)$ we have not attempted to match our parameters to Naaman's experiments, and have simulated simple scattering problems with two (rather than many) electronic states -- more on this below.
For these reasons, one cannot draw any meaningful conclusions yet regarding whether or not a nuclear Berry force is important for CISS; and yet, it is clear that future research should focus on this point.
From our perspective, we intend to run semi-classical surface hopping calculations in the future using a recently defined FSSH protocol for treating complex Hamiltonians.\cite{Miao2019}

%And so, in the future, the next steps forward must address:
%(1) The Berry phase effect in real systems, especially in CISS systems. How the Berry phase effect participate in spin selectivity and how the strength compared to pure electronic effect are of our great interest. In order to investigate larger and complicated systems, semi-classical methods need to be developed, such as FSSH.
%(2) Quantitative theory of Berry phase effect. Even the spin polarization is shown proportional to Berry curvature, its quantified relation with system geometry is not clear. In our simulations, spin polarization have already shown a various dependence on system geometry (both the potential energy surface and the phase), so in a real system with much more complicated geometry, it is not straightforward to predict the actual effect.

\subsection{The question of multiple electronic states} \label{sec:discussionmultiple}

The third and final outstanding question regarding Berry force effects on reaction is the question of multiple electronic states. As we demonstrated in Sec. \ref{sec:twoterminal}, the two-terminal scattering case is special in so far as the fact that, when only two terminals are open, the overall transmission from \#1 to \#2 {\em must} be symmetric and equal the overall transmission from \#2 to \#1 {\em even in the presence of an external magnetic field that breaks TRS}. Obviously, this is a quirk of the two-terminal case. Consider the following diagram with three symmetric terminals:
\begin{figure}[H]
	\begin{center}
		\includegraphics[width=0.3\columnwidth]{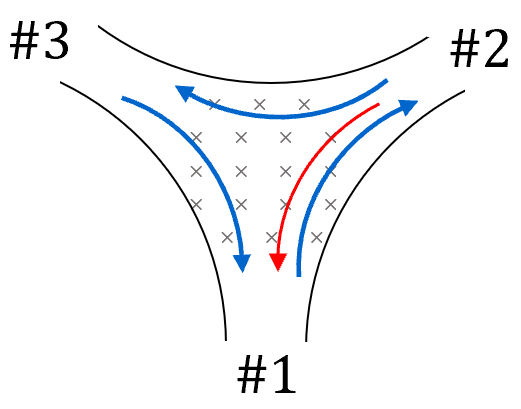}
	\end{center}
	\caption{Schematic view of a symmetric three-terminal system with external magnetic field (pointing inward, marked as gray crossings). Assuming the trajectories of particles are bent to right, the blue and red arrow show the favored flow and unfavored flow. In such system, the rate constants $k_{1\rarr 2}=k_{2\rarr 3}=k_{3\rarr 1}$, but in general $k_{1\rarr 2} \ne k_{2\rarr 1}$.}
	\label{fig:threeterm}
\end{figure}
\noindent Clearly, with a magnetic field on, we will not find that $k_{1\rarr 2} = k_{2\rarr 1}$. Instead, by symmetry, we will find that $k_{1\rarr 2}=k_{2\rarr 3}=k_{3\rarr 1}$. Thus, the impact of Berry's phase must be examined carefully with many electronic states (and more than two terminals).

Finally, there is one more key point worth mentioning. For the two-state system considered above, a Berry force emerged from the dependence of the phase of the diabatic coupling $Wu(x,y)$ (see Eqs.~\eqref{eq:parambent1},\eqref{eq:parambent2},\eqref{eq:paramlinear1},\eqref{eq:paramlinear2}). And in this regard, one might question whether such a non-Condon term can have a large effect on the overall spin-dynamics of a given system.
After all, non-Condon terms are entirely absent from the standard Marcus theory. That being said, consider a case of three electronic states (e.g. corresponding to super-exchange) where two diabatic states come close together but are also coupled weakly to a third state:
\begin{align} \label{eq:threestateh}
    H_{el}(\bm{R}) = \begin{bmatrix}E_1(\bm{R}) & V_{12} & V_{13} \\ V_{12}^* & E_2(\bm{R}) & V_{23} \\ V_{13}^* & V_{23}^* & E_3(\bm{R}) \end{bmatrix}
\end{align}

Notice that, if $E_1(\bm{R})$, $E_2(\bm{R})$ and $E_3(\bm{R})$ are not parallel, we will still find a nonzero Berry force if the couplings $V_{12}$, $V_{13}$ and $V_{23}$ are all nonzero and not all of them are real. Moreover, if $E_1(\bm{R})$ and $E_2(\bm{R})$ are close together, the Berry force can be very large. Thus, in the future, in order to model Berry force effects on nonadiabatic dynamics, it will be crucial to investigate systems with more than two (spatial) electronic states.

%\subsection {FSSH Results}

%To investigate whether FSSH is able to capture the Berry's phase effect correct quantitatively, we simulate the two-state system described in Section III using FSSH, with the velocity rescaling scheme describled in Appendix C. Shown in Fig. \ref{figure:fssh}, the transmission $T(E)$ given by FSSH is slightly higher than quantum results, but the transmission bias $\Delta T(E)$ and polarization $P(E)$ has overall good agreement when the energy is out of the peak region and when the system is more adiabatic, i.e. the coupling $C$ is larger. When the energy is close to barrier height (0.01 here), FSSH results show large deviation from quantum results. From these results, we conclude that FSSH is capable of capturing the spin polarization induced by Berry curvature.

%\begin{figure}[H]
%	\begin{center}
%		\includegraphics[width=\columnwidth]{figfssh.eps}
%	\end{center}
%	\caption{Comparison of the total transmission $T(E)$ (left), transmission rate difference $\Delta T(E)$ (center) and polarization $P(E)$ (right) given by quantum scattering and FSSH of two-state system. The parameters are $A=0.02$, $W=0.2$ and $C=0.01$ (up) or $0.005$ (down). Other parameters are same as quantum calculation.}
%	\label{figure:fssh}
%\end{figure}

\section{Conclusion}

In summary, we have investigated the effect of Berry's phase (or more importantly, Berry force) on nuclear dynamics in the context of a complex model Hamiltonian with two electronic states.
Our quantum scattering simulations have demonstrated that magnified spin polarization can arise for nuclear geometries of a ``chiral nature'' (or really just nuclear geometries without inversion symmetry) and our scattering results can even be roughly predicted using an analytic expression for Berry force at the crossing point.
Overall, our findings may provide some insights into recent CISS experiments, suggesting a possible role for nuclear motion in promoting spin selective electronic transport, though a great deal more research (likely with large numerical simulations) will be needed to confirm such a hypothesis.

\begin{acknowledgement}
This material is based on the work supported by the National Science Foundation under Grant No. CHE-1764365. J.E.S. acknowledges support from a Camille and Henry Dreyfus Teacher Scholar Award. J.E.S. thanks Abraham Nitzan for very stimulating conversations.
\end{acknowledgement}

%\section*{Data Availability}
%The data that support the findings of this study are available from the corresponding author upon reasonable request.

\appendix
\section{Derivation of The Effective Magnetic Field} \label{sec:appendixbfield}

Here we derive the effect magnetic field for the two-state system with bent geometry (Eq.~\eqref{eq:btwostatebent}).
In the two-state system, the electronic Hamiltonian for the up spin is given by
\begin{align}
    H = \begin{bmatrix} E_1(\bm{R}) & V_t(\bm{R}) \\ V_t^*(\bm{R}) & E_2(\bm{R}) \end{bmatrix}
\end{align}
where $E_1,E_2,V_t$ are given in Eqs.~\eqref{eq:param1_1},\eqref{eq:param1_2},\eqref{eq:parambent1}. Now, we define $\tilde{\theta} = \epsilon(\theta-\frac{\pi}{4})$, $\tilde{A}(\theta)=A \tanh{\tilde{\theta}}$, $\tilde{C}(\theta)=Ce^{-\tilde{\theta}^2}$, so that
\begin{align}
    H = \begin{bmatrix} A+\tilde{A} && \tilde{C}e^{iWu} \\ \tilde{C}e^{-iWu} && A-\tilde{A}\end{bmatrix}
\end{align}
The two adiabatic states and energies are
\begin{align}
    \psi_0 = \frac{1}{\eta_0}\begin{bmatrix}\tilde{A}-\sqrt{\tilde{A}^2+\tilde{C}^2} \\ \tilde{C}e^{-iWu}\end{bmatrix} \\
    \psi_1 = \frac{1}{\eta_1}\begin{bmatrix}\tilde{A}+\sqrt{\tilde{A}^2+\tilde{C}^2} \\ \tilde{C}e^{-iWu}\end{bmatrix} \\
    E_0^a = A-\sqrt{\tilde{A}^2+\tilde{C}^2} \\
    E_1^a = A+\sqrt{\tilde{A}^2+\tilde{C}^2}
\end{align}
where $\eta_0 = \sqrt{{(\tilde{A}-\sqrt{\tilde{A}^2+\tilde{C}^2})}^2+\tilde{C}^2}$, $\eta_1 = \sqrt{{(\tilde{A}+\sqrt{\tilde{A}^2+\tilde{C}^2})}^2+\tilde{C}^2}$ are the normalization constants.
By substituting the equations above into Eq.~\eqref{eq:drvcoupling}, we recover
\begin{align}
    \bm{d}_{01} &= \frac{\mel{\psi_0}{\nabla H}{\psi_1}}{E_1^a-E_0^a} \nonumber \\ 
    &= \frac{1}{\eta_0\eta_1}\Big(\frac{\tilde{A}\tilde{C}\nabla\tilde{C} - \tilde{C}^2\nabla\tilde{A}}{\sqrt{\tilde{A}^2+\tilde{C}^2}} - iW\tilde{C}^2\nabla u(x,y)\Big) \nonumber \\
    &= -\frac{1}{\eta_0\eta_1}\Big(
        \frac{\tilde{C}^2}{\sqrt{\tilde{A}^2+\tilde{C}^2}}\frac{\epsilon}{R}\Big(
            A\sech^2{\tilde{\theta}} + 2A\tilde{\theta}\tanh{\tilde{\theta}}\Big)\bm{e}_{\theta}
            + iW\tilde{C}^2\nabla u(x,y) \Big) \label{eq:exprdrvcoupling}
\end{align}
Let us define $\nabla u(x,y) = \gamma(x,y)\bm{e}_R + \zeta(x,y)\bm{e}_\theta$. Then
\begin{align} \label{eq:exprdrvcoupling2}
    \bm{d}_{01} = -\frac{1}{\eta_0\eta_1}\Big(
            \frac{\tilde{C}^2}{\sqrt{\tilde{A}^2+\tilde{C}^2}}\frac{\epsilon}{R}\Big(
                A\sech^2{\tilde{\theta}} + 2A\tilde{\theta}\tanh{\tilde{\theta}}\Big)\bm{e}_{\theta}
                + iW\tilde{C}^2\zeta(x,y)\bm{e}_\theta + iW\tilde{C}^2\gamma(x,y)\bm{e}_R \Big)
\end{align}
Because the component in the $\bm{e}_R$ direction (proportional to $\gamma(x,y)$) is exclusively imaginary, the real component in the $\bm{e}_\theta$ direction (proportional to $\zeta(x,y)$) will not contribute to the final Berry force.
Substituting Eq.~\eqref{eq:exprdrvcoupling2} into Eq.~\eqref{eq:bfield} gives the effective magnetic field:
\begin{align}
    \bm{B}_0 &= -i\bm{d}_{01}\times\bm{d}_{10} = i\bm{d}_{01}\times\bm{d}_{01}^* \nonumber \\
    &= -\frac{1}{{(\eta_0\eta_1)}^2} \frac{2WA\tilde{C}^3\epsilon}{R}\frac{\tilde{C}}{\sqrt{\tilde{A}^2+\tilde{C}^2}}(\sech^2{\tilde{\theta}}+2\tilde{\theta}\tanh{\tilde{\theta}})\gamma(x,y)\bm{e}_z \nonumber \\
    &=  -\frac{W\epsilon}{2R}\frac{A\tilde{C}^2}{{(\tilde{A}^2+\tilde{C}^2)}^{3/2}}(\sech^2{\tilde{\theta}}+2\tilde{\theta}\tanh{\tilde{\theta}})\gamma(x,y)\bm{e}_z
\end{align}
Finally, using the fact that $\tilde{A} = \frac{E_1 - E_2}{2}$ and $\tilde{C}^2=\abs{V_t}^2$, we arrive at Eq.~\eqref{eq:btwostatebent}.

\bibliography{ComplexJ,footnotes} 

\begin{mcitethebibliography}{101}
\bibitem[foo()]{footnote1}
For the above Hamiltonians, a different expression can also be generated along
  the excited state, but is not provided here.\relax
\mciteBstWouldAddEndPunctfalse
\mciteSetBstMidEndSepPunct{\mcitedefaultmidpunct}
{}{\mcitedefaultseppunct}\relax
\EndOfBibitem
\bibitem[foo()]{footnote2}
One could perhaps resolve spin currents if one attempted to split apart the
  outgoing wavepackets into multiple sub-terminals, and resolve how the product
  wavepacket varied in the $x$ direction after the first crossing.\relax
\mciteBstWouldAddEndPunctfalse
\mciteSetBstMidEndSepPunct{\mcitedefaultmidpunct}
{}{\mcitedefaultseppunct}\relax
\EndOfBibitem
\bibitem[foo()]{footnote3}
This can be easily checked by allowing for a magnetic field and observing that
  the stationary (Boltzmann) solution to the Fokker-Planck equation is
  unchanged.\relax
\mciteBstWouldAddEndPunctfalse
\mciteSetBstMidEndSepPunct{\mcitedefaultmidpunct}
{}{\mcitedefaultseppunct}\relax
\EndOfBibitem
\end{mcitethebibliography}
\bibliographystyle{apsrev}
	
\end{document}